\documentclass[%
 reprint,
superscriptaddress,
 amsmath,amssymb,
 aps,
]{revtex4-2}

\usepackage{graphicx}
\usepackage{dcolumn}
\usepackage{bm}
\usepackage{hyperref}
\usepackage{color}

\usepackage[caption=false,listofformat=subsimple,labelformat=simple]{subfig}

\begin{document}

\preprint{APS/123-QED}

\title{Effective viscosity and elasticity in dense suspensions under impact: Toward a modeling of walking on suspensions}

\author{Pradipto}
\email{fu2949@go.tuat.ac.jp}
\affiliation{%
Department of Mechanical Systems Engineering, Tokyo University of Agriculture and Technology, Koganei, Tokyo 184-8588, Japan \footnote{Present address}
}%
\affiliation{%
 Yukawa Institute for Theoretical Physics, Kyoto University, Kitashirakawa Oiwake-Cho, Sakyo-ku, Kyoto 606-8502, Japan
}%

\author{Hisao Hayakawa}
\affiliation{%
 Yukawa Institute for Theoretical Physics, Kyoto University, Kitashirakawa Oiwake-Cho, Sakyo-ku, Kyoto 606-8502, Japan
}%

\date{\today}
\begin{abstract}
The elastic response of dense suspensions under an impact is studied using coupled Lattice Boltzmann Method and Discrete Element Method (LBM-DEM) and its reduced model.
We succeed to extract the elastic force acting on the impactor in dense suspensions, which can exist even in the absence of percolating clusters of suspended particles.
We then propose a reduced model to describe the motion of the impactor and demonstrate its relevancy through the comparison of the solution of the reduced model and that of LBM-DEM. 
Furthermore, we illustrate that the perturbation analysis of the reduced model captures the short-time behavior of the impactor motion quantitatively.
We apply this reduced model to the impact of a foot-spring-body system on a dense suspension, which is the minimal model to realize walking on the suspension.
Due to the spring force of the system and the stiffness of the suspension, the foot undergoes multiple bounces.
We also study the parameter dependencies of the hopping motion and find that multiple bounces are suppressed as the spring stiffness increases.
\end{abstract}

\maketitle

\section{Introduction}

The phenomenon of being able to walk on suspensions has attracted the interest of both scientists and the general public \cite{brown2014, ness2022}.
Such impact-induced hardening of dense suspensions is often chosen as an example of discontinuous shear thickening (DST) \cite{brown2014}, but it has already been shown that the underlying mechanism of impact-induced hardening is different from that of DST \cite{pradipto2021}.
In fact, impact-induced hardening is a transient process in which only normal stress becomes large and the system is heterogeneous, whereas DST is a steady process in which both shear and normal stresses become large and the system is homogeneous. 

Most physical studies of impact-induced hardening use a free-falling impactor or a constant-velocity penetrating intruder. 
Using a free-falling impactor, Ref. \cite{waitukaitis2012} reported the existence of a localized rigid region under the impactor, called the dynamically jammed region (DJR).
As such a DJR grows in size, Ref. \cite{waitukaitis2012} proposed the added-mass model, which treats the impact as an inelastic collision between the impactor and the DJR.
Then, Ref. \cite{han2015} visualized the flow field in the dense suspension around the penetrating intruder, and found that the strain rate peaked on the boundary of the DJR.
Inspired by this observation, Ref. \cite{brassard2020} proposed a model based on the viscous force acting on the boundary of the DJR.
However, none of the above models can explain the existence of elastic response of dense suspensions under impacts such as fracture \cite{roche2013}, high stress near the boundary \cite{maharjan2018}, and rebound of the impactor \cite{egawa2019}.
In Ref. \cite{maharjan2018}, a constitutive model was proposed and the modulus of elasticity was measured when the DJR spans from the impactor to the boundary.
Then, the viscoelastic response of dense suspensions under an impact is captured using the floating + force chains model \cite{pradipto2021b}, where the elastic force is only finite when the force chains of contacting suspended particles percolate from the impactor to the bottom boundary.
However, such an analysis requires data on the position of the suspended particles to resolve the force chains and calculate the number of percolated force chains.
Moreover, the prediction of the floating + force chain model that percolating force chains are needed to get elastic response is questionable, because this denies the possibility of elastic response of suspensions confined in a deep container.

The motion of a running or walking person on a suspension liquid is more complicated than that of a free-falling impactor or a penetrating intruder.
An approach to study the walking motion on the suspensions was described in Ref. \cite{mukhopadhyay2018}.
They discussed the maximum penetration depth of a foot for different impact velocities corresponding to walking, jogging, and running \cite{maharjan2018}.
They also showed that the added mass model is not sufficient to recover the response of the suspensions under running motion.
Some studies adopted mechanical models for the locomotion of legged animals.
One of the simplest models is the spring-mass model inspired by biomechanical observations \cite{Blickchan1989}.
In the spring-mass model, the human leg is represented by a spring, and the human body is simply represented by a mass point.
Such a model has been realized as a one-legged hopping robot \cite{raibert1986}.
Thus, the realization of hopping, i.e. multiple bounces after the rebound is crucial to describe walking or running on a liquid.
However, little is known about the dynamics of multiple bounces after an impact on dense suspensions.

Based on the current situation of related studies, we have two motivations for this study.
The first motivation is to clarify the role of elasticity in dense suspensions, and whether such elasticity can exist even in the absence of percolating clusters of suspended particles.
Then, we propose a reduced equation of motion for the impactor, which is sufficiently correct to reproduce the motion of the impactor by a full set of equations of motions of the impactor and grains including the hydrodynamic interactions among grains and rotations of grains. 
We also verify the existence of elastic force acting on an impactor even in the absence of percolating clusters of suspended particles.
Our second motivation is to extend the motion of a single impactor to the motion of a body with internal degrees of freedom because hopping is not captured by previous known models (e.g., added-mass model or viscous model), nor by the model of the impactor without internal degrees of freedom used in our previous studies in Refs. \cite{pradipto2021,pradipto2021b}. 
Inspired by the previous models used in Refs. \cite{Blickchan1989,raibert1986}, this paper studies the motion of a foot-spring-body system coupled with the LBM-DEM model introduced in Refs. \cite{pradipto2021,pradipto2021b} on dense suspensions to realize, at least, the hopping of the body on the suspension fluid. 
We also apply the reduced model to the foot-spring-body system and verify that the reduced model captures the bouncing dynamics on the suspension.

This paper is organized as follows.
In Sec. \ref{SecII}, we explain our simulation setup and evaluate the viscosity and elastic force acting on the impactor using the coarse-grained technique during the impact process.
Then, we propose an empirical law for the elastic force. 
We illustrate that a perturbation theory in which the linear correction to the floating model \cite{pradipto2021b} is involved gives us a quantitatively correct result for the short-time behavior of the impactor.
In Sec. \ref{SecIII}, we describe the simulation setup for the foot-spring-body model and present the hopping motion of such a system in order to clarify the criterion for the hopping motion.
In Sec. \ref{Conclusion}, we summarize our results and discuss the future prospects of this study.
In Appendix \ref{app:lbm}, we describe the details of the coupled Lattice Boltzmann Method and Discrete Element Method (LBM-DEM) used in our simulation.
In Appendix \ref{app:exp}, we compare our simulation results with relevant experiments.
In Appendix \ref{app:surf}, we describe the details of performing integrals on the impactor surface.
Finally, in Appendix \ref{app:perturbation}, we present the details of the perturbation approach.

\section{Evaluation of viscosity and elastic force around the impactor}\label{SecII}

In this section, we analyze a free-falling impactor on a dense suspension.
This section consists of five subsections.
In Sec. \ref{SecIIA}, we briefly explain the setup for a free-falling impactor simulation. 
In Sec. \ref{SecIIB}, we derive a reduced equation of motion for the impactor.
In Sec. \ref{SecIIC}, we explain the technique to describe the local fields such as the stress field, strain field, and strain rate field using a coarse-grained method.
This enables us to evaluate the force acting on the impactor.
In Sec. \ref{SecIID}, we evaluate the elastic force acting on the impactor, and propose an empirical expression of the elastic force.
In Sec. \ref{SecIIE}, inserting the obtained results into the reduced equation of motion for the impactor, we obtain the motion of the impactor, which recovers the results of a full set of equations of LBM-DEM. 

\subsection{Setup for a free-falling impactor simulation}\label{SecIIA}

\begin{figure}[htbp]
    \centering
		\includegraphics[width=0.7\linewidth]{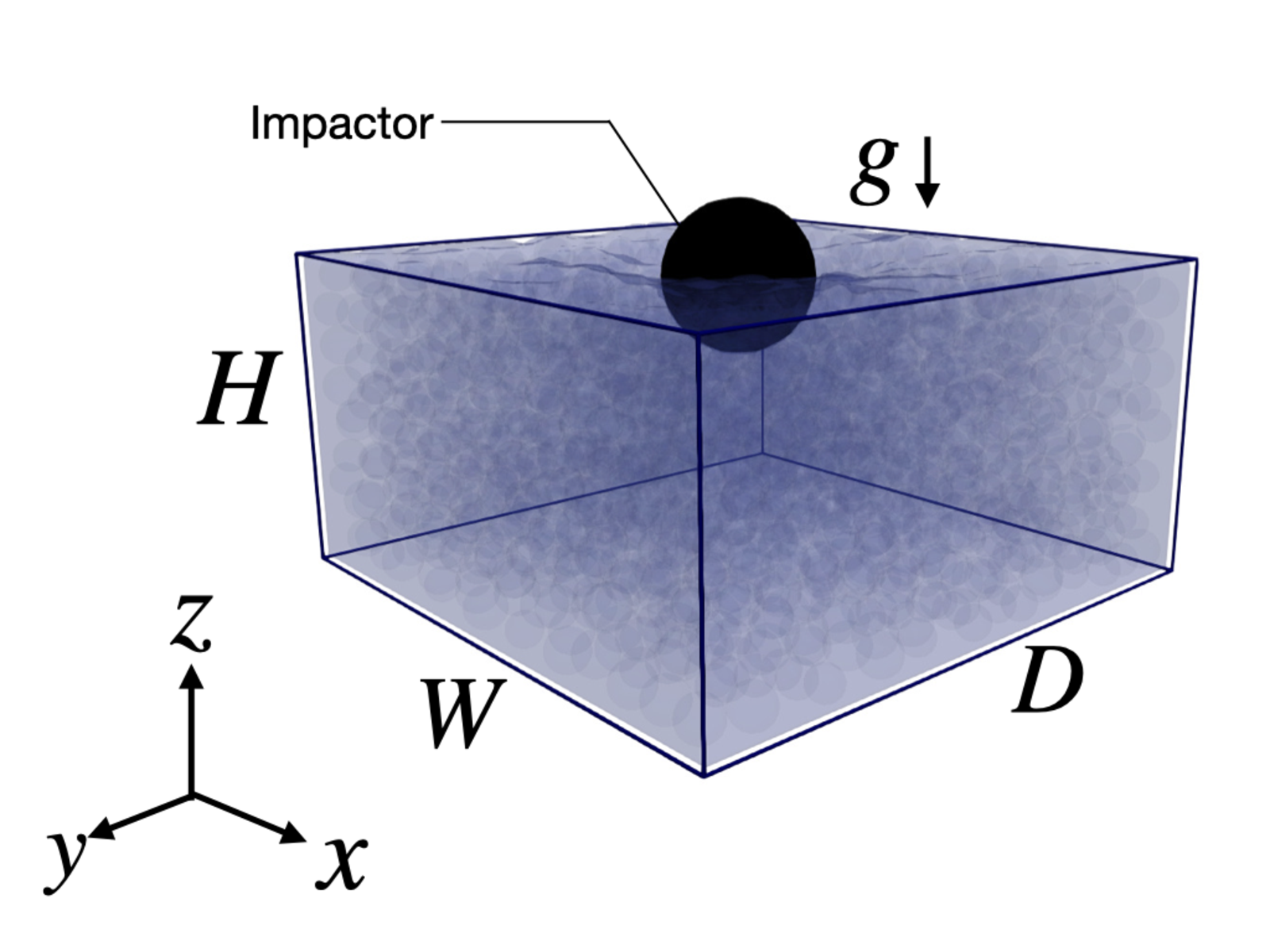}
	\caption{Illustration of our simulation setup.}
	\label{fig:1}
\end{figure}

\begin{figure*}[htbp]
    \centering
	\subfloat[]{\label{fig:2a}%
		\includegraphics[width=0.3\linewidth]{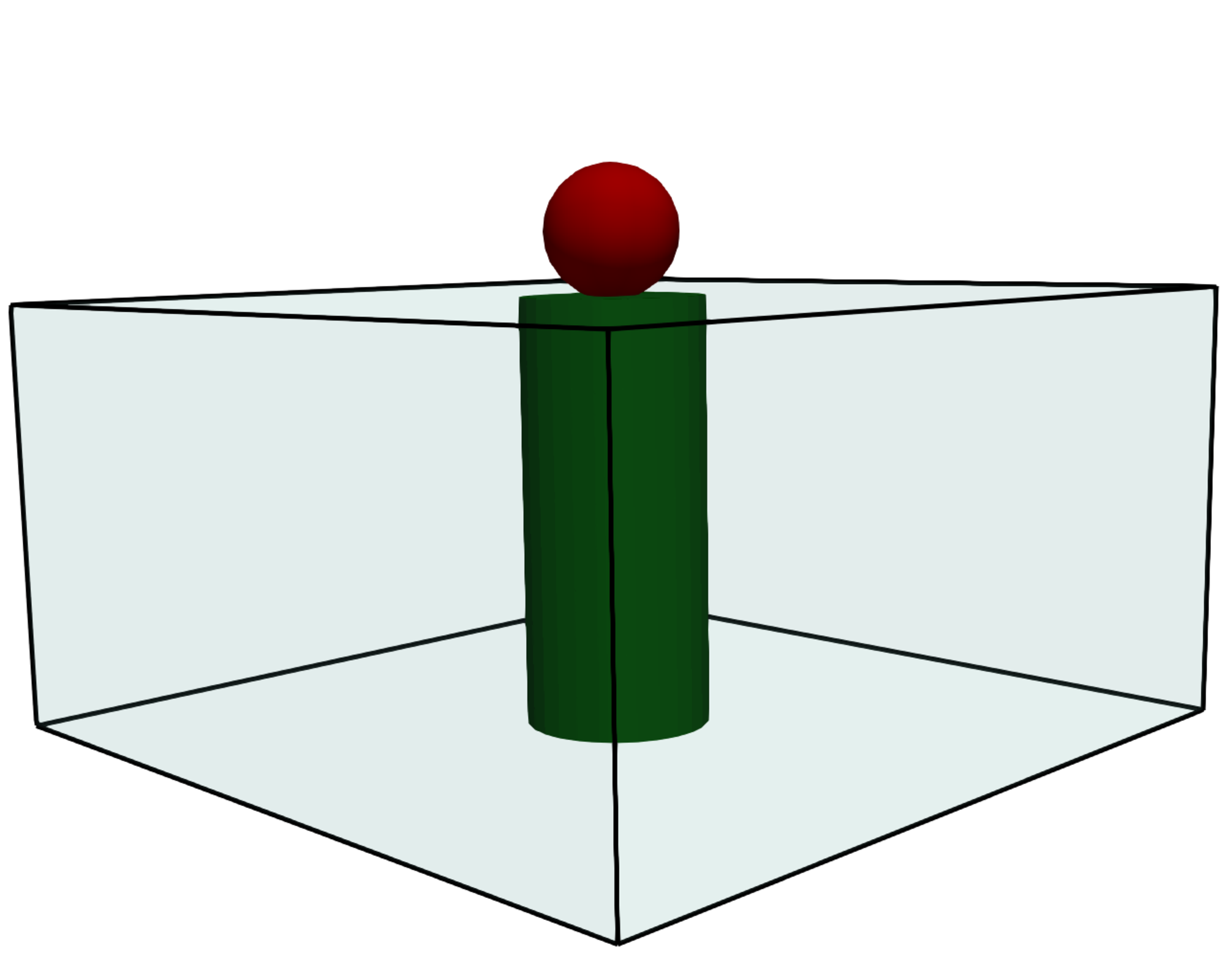}
	}
		\subfloat[]{\label{fig:2b}%
		\includegraphics[width=0.3\linewidth]{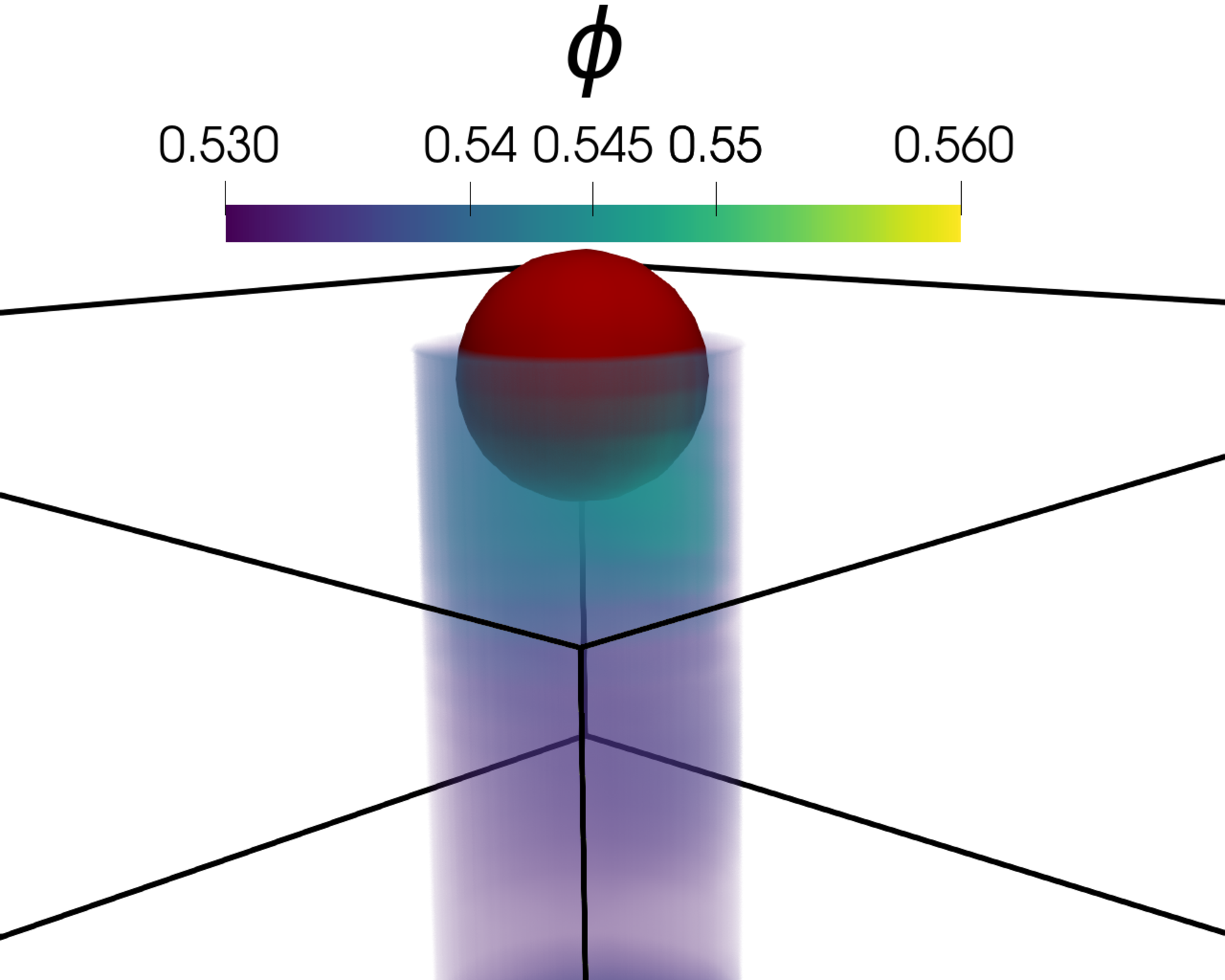}
	}
 		\subfloat[]{\label{fig:2c}%
		\includegraphics[width=0.3\linewidth]{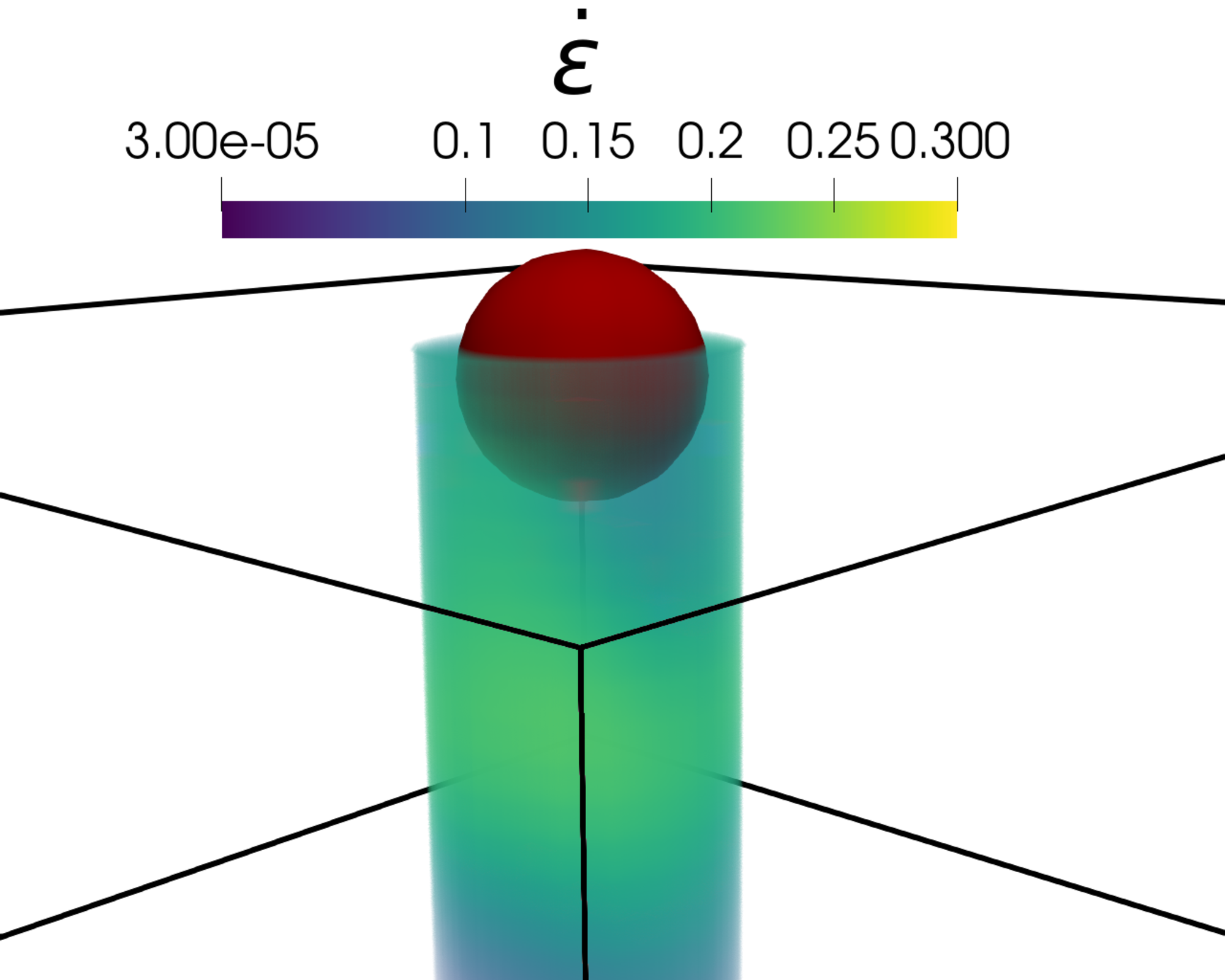}
	}

  		\subfloat[]{\label{fig:2d}%
		\includegraphics[width=0.3\linewidth]{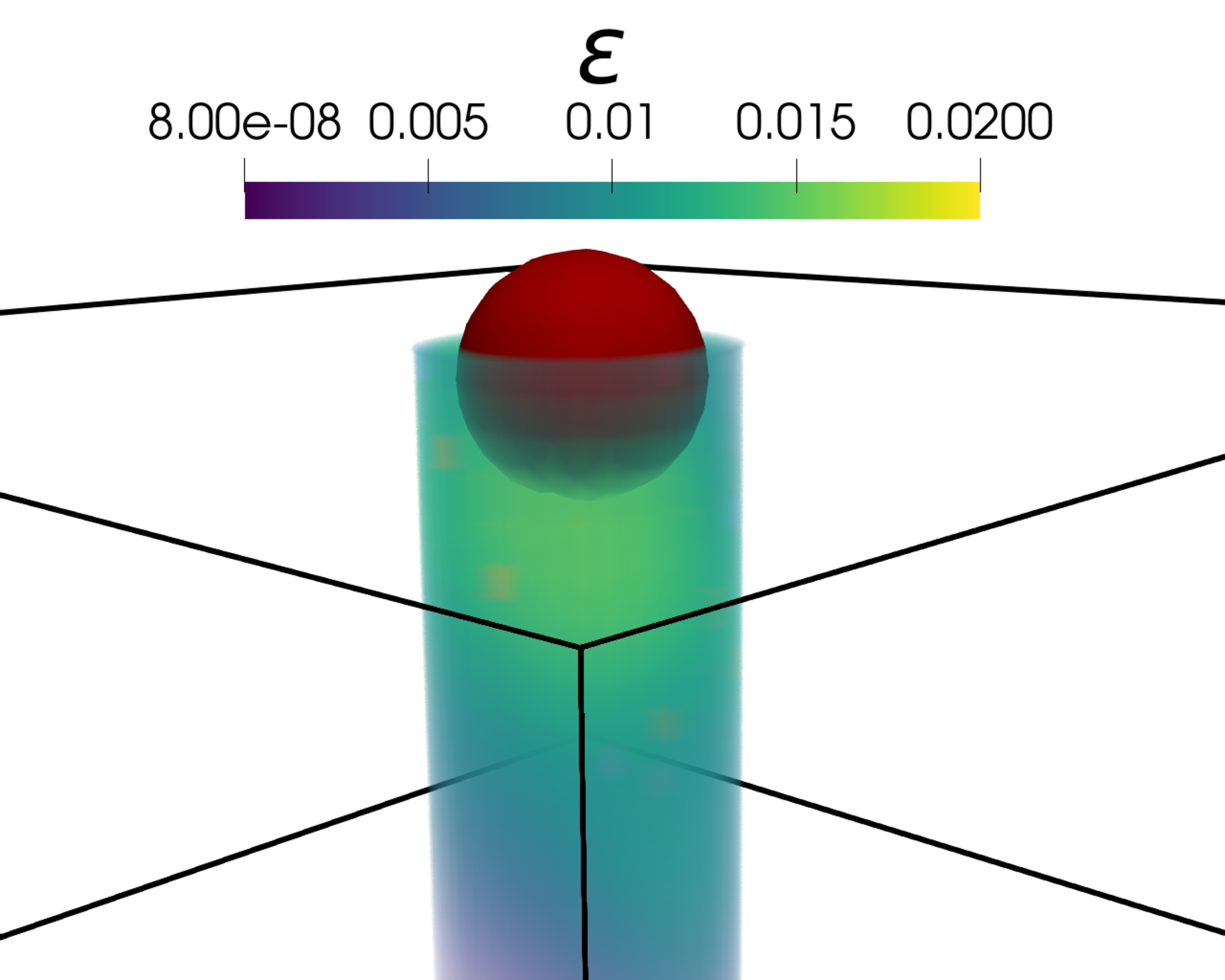}
	}
  		\subfloat[]{\label{fig:2e}%
		\includegraphics[width=0.3\linewidth]{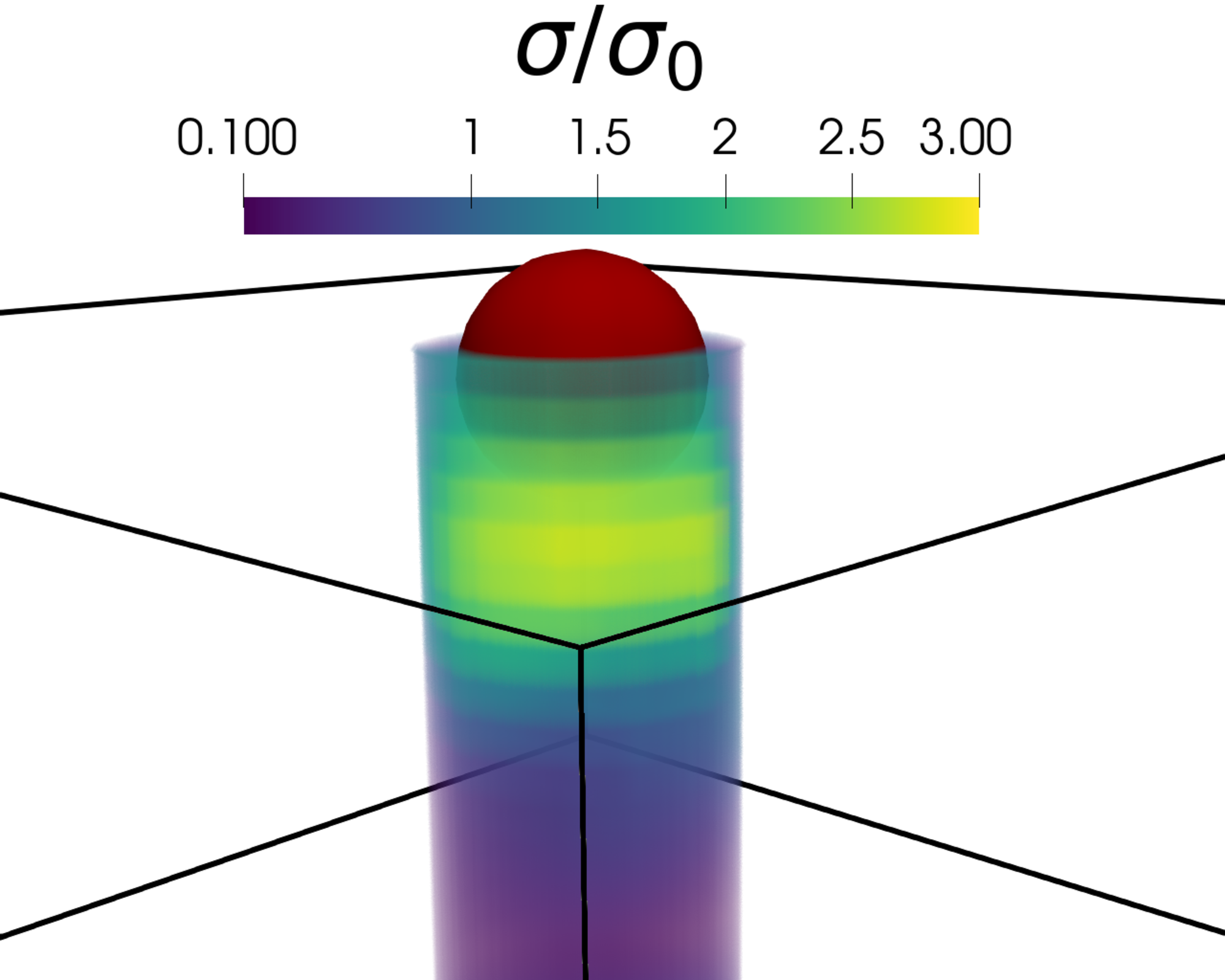}
	}
	\caption{Visualizations of coarse-grained variables for $\phi = 0.53$, $u_0 = 4.5 u^{*}$, $t = 0.06 t_g$, where we present (a) selected region for visualization (green cylinder), (b) local volume fraction $\phi$, (c) local scalar strain rate $\dot{\varepsilon}$, (d) local scalar strain $\varepsilon$, and (e) local stress $\sigma$, respectively. 
 }
	\label{fig:2}
\end{figure*}

Let us consider an impactor falling into a suspension (see Fig. \ref{fig:1}), where $z$ denotes the vertical direction and the gravity acts along the negative $z$ direction. 
Through this paper, we set $z=0$ on the surface of the suspension before the collision of the impactor.
This means that inside the suspension, $z<0$ is always satisfied.

As a basic set of equations, we adopt the coupled LBM-DEM model as in Refs \cite{pradipto2021,pradipto2021b}.
This model assumes that the fluid flow can be described by the Stokes equation.
This means that the diameters of the suspended particles are of the order of $100 \mu \rm m$.
This model includes equations of motion for the impactor and suspended particles, where the forces acting on the impactor and suspended particles include the contact force, hydrodynamic interactions, lubrication, electrostatic repulsive interactions, and gravity, as well as the torque balance equations for the impactor and suspended particles as shown in Appendix \ref{app:lbm}.
The simulation setup is as follows \cite{pradipto2021b}.
We analyze a mixture of $N$ suspended particles with a volume fraction $\phi_0:= (2 \pi/3)N(a_{\rm min}^3+a_{\rm max}^3)/V_{\rm box}$, where $V_{\rm box}$ is the volume of the container, $V_{\rm box}:=W \times H \times D$ with the width $W$, the height $H$, and the depth $D$, and $a_{\rm min}$ and $a_{\rm max}$ are the radii of the smaller and larger suspended spheres, respectively (see Fig. \ref{fig:1}).
Here, we adopt $a_{\rm max} = 1.2 a_{\rm min}$ to avoid crystallization in high density.
We analyze only the case where the number of smaller spheres is equal to the number of larger spheres.
Throughout this paper, we assume perfect density matching between the solvent and the suspended particles, where the densities of the particles $\rho_p$ and the solvent $\rho_f$ satisfy the relation $\rho_p=\rho_f$.
In this section, a spherical impactor with diameter $D_I$ (radius $a_I$) and density $\rho_I$ is released from height $H_0$, which corresponds to the impact velocity $u_{0} = \sqrt{2 g H_0}$ with gravitational acceleration $g$.  
In our simulation, $\rho_I$ and $D_I$ satisfy $\rho_I= 4 \rho_f$ and $D_I=6 a_{\rm min}$, respectively.
We also introduce the time scale $t_g = \sqrt{a_I/2g}$, the velocity scale $u^{*} = \sqrt{2 g a_I}$,the force scale $F_g=\frac{4}{3} \pi \rho_f a_{I}^{3} g$, and stress scale $\sigma_0 = F_g / a_I^{2}$.
For most of the cases considered in this paper, we use $\phi_0=0.53$, $H=3D_I$, $W=D=6D_I$, and $N=2200$.

In Appendix \ref{app:exp}, we show that a full set of equations based on the LBM-DEM model can reproduce the experimental results~\cite{egawa2019}, although the dimensionless time in experimental data is not scaled by $t_g$, but by a different time scale. 
This discrepancy in the time scale between the experiment and simulation may originate from the finite size effect as indicated in Ref.~\cite{pradipto2021}.
It should be noted that we cannot get any physical insight into the motion of the impactor by simulating a full set of equations because the simulation is expensive and we need to know the motions of grains in suspensions.
Instead, if we can obtain an equation of the motion of the impactor without referring to the motion of suspended particles, its advantage is obvious because such an equation can be easily solved and the analytical expression of the motion of the impactor can be used in some limited situations as in Ref.~\cite{pradipto2021b}.
Since various useful results have already been obtained based on such an approach in the previous studies \cite{pradipto2021,waitukaitis2012,brassard2020,egawa2019,pradipto2021b}, we also adopt a reduced equation of motion of the impactor in this paper.

\subsection{Equation of motion of impactor}\label{SecIIB}
\begin{figure*}[htbp]
    \centering
		\subfloat[]{\label{fig:3b}%
		\includegraphics[width=0.4\linewidth]{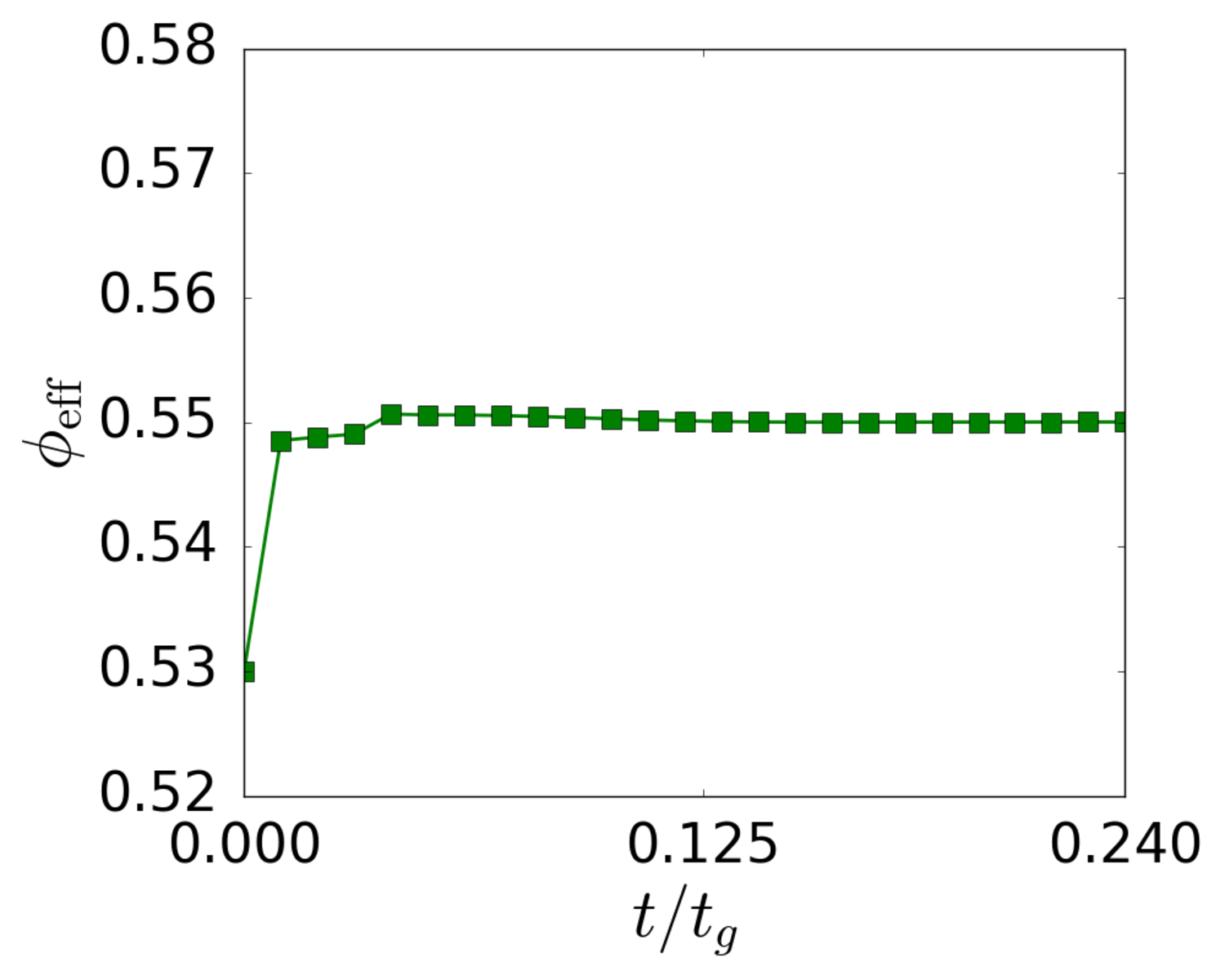}
	}
 		\subfloat[]{\label{fig:3c}%
		\includegraphics[width=0.4\linewidth]{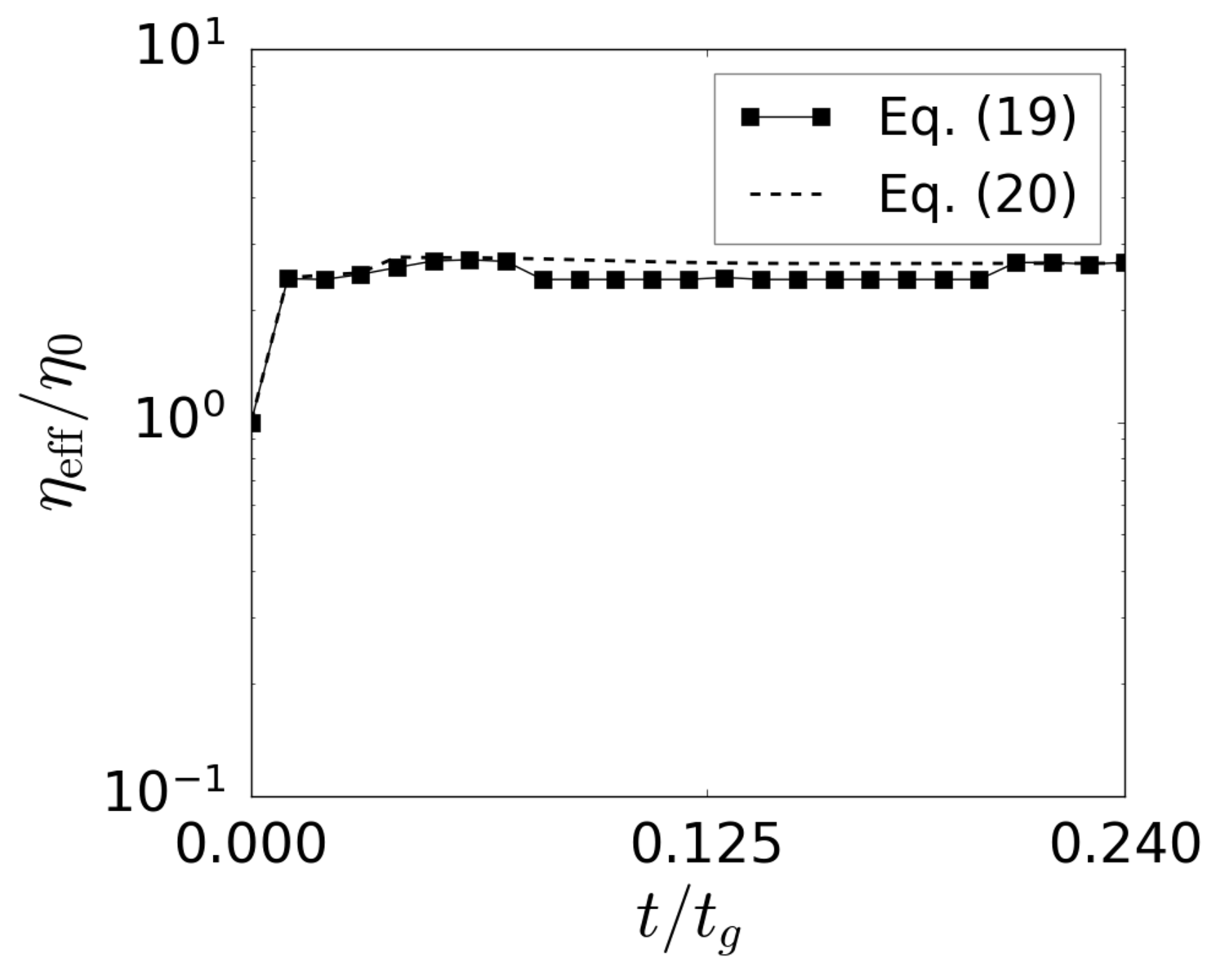}
	}
	\caption{(a) A plot of the time evolution of effective volume fraction $\phi_{\rm eff}$ on $S$ for $\phi=0.53$ and $u_0=4.5u^{*}$. (b) A plot of the time evolution of effective viscosity $\eta_{\rm eff}$ on $S$ for $\phi=0.53$ and $u_0=4.5u^{*}$.
 }
	\label{fig:3}
\end{figure*}
In the reduced model, the equation of motion for a free-falling impactor along $z-$direction for the lowest point of the impactor $z:=z_I-a_I$, where $z_I$ is the vertical position of the center of mass of the impactor, onto dense suspensions can be written as
\begin{equation}
    m_I \ddot{z} = - m_I \tilde{g} - 3 \pi \eta_{\rm eff} \dot{z} |z| - F_{\rm el} (z),
    \label{eq:eom}
\end{equation}
where $m_I$ is the mass of the impactor, and $\tilde{g}$ is the effective gravity acceleration defined as $\tilde{g}: = g (\rho_I-\rho_f)/\rho_f$ with the densities of the impactor$\rho_I$ and the solvent $\rho_f$.
The second term on the right-hand side (RHS) of Eq. \eqref{eq:eom} has been introduced in Ref. \cite{brassard2020}, and its validity has been verified in Ref. \cite{pradipto2021b}.
In order to propose a model that does not need to refer to the simulation data, we need to determine two unknown variables: (i) the effective viscosity $\eta_{\rm eff}$ and (ii) the elastic force $F_{\rm el}$.
The previous studies~\cite{brassard2020,pradipto2021b} suggest that $F_{\rm el}$ is not important in the early stage of the impact.

Before the explanation of the method of how to obtain $F_{\rm el} (z)$ om Eq.~\eqref{eq:eom}, we briefly comment on the gravity acceleration $\tilde{g}$ and the surface deformation of the suspension after the impact.
First, $\tilde{g}$ in Eq.~\eqref{eq:eom} is an over-simplified treatment, because this treatment is correct if the impactor is completely inside the suspension but is not correct if the impactor is partially inside the suspension.
This means that the time scale of our simulation might be different from that in real experiments.
Nevertheless, we have already verified that such a simplification gives us a reasonable result as shown in Ref.~\cite{pradipto2021b}.
Thus, we adopt this over-simplified model.
Second, the surface of the suspension liquid in the LBM-DEM simulations is deformed after the impact as in real experiments, and thus, the actual surface can be higher or lower than $z=0$ \cite{pradipto2021,pradipto2021b}.
However, since the ripple on the surface does not contribute to the force acting on the impactor, such deformation of the suspension surface is ignored in our coarse-grained procedure.

\subsection{Local variables using coarse-grained method}\label{SecIIC}

In this subsection, we describe the method for obtaining local variables within suspensions that are relevant for elucidating the behavior of viscosity and elastic force on the impactor during the impact process.
The variables we use are (i) local volume fraction, (ii) local strain rate, (iii) local strain, and (iv) local stress.
The approximate description of such continuum fields from discrete particle data can be carried out using the coarse-grained method, which has been used for granular materials \cite{zhang2010,saitoh2013}.
Here, all variables within suspensions are calculated on a rectangular grid with a lattice constant $0.5 a_{\rm min}$.

The local volume fraction $\phi (\bm{r})$ can be expressed with a coarse-grained function $\Phi(\bm{r})$ as
\begin{equation}
    \phi (\bm{r},t) := \sum_i \Phi (\bm{r} - \bm{r}_i (t)),
    \label{eq:cg_rho}
\end{equation}
where $\bm{r}$ and $\bm{r}_i$ are the field position and the position of $i-$th particle, respectively.
Here, we adopt
\begin{equation}
    \Phi(\bm{r} - \bm{r}_i) := \frac{1}{(w\sqrt{2 \pi})^3} \exp\left[ -\frac{(\bm{r}-\bm{r}_i)^2}{2w^2}\right] .
\end{equation}
For all the results presented here, we adopt a width of $w=6a_{\rm min}$. 
To satisfy the boundary conditions on the wall, mirrored copies of the particle configurations are required on each side of the wall before applying the coarse-grained methods \cite{ries2014}.
In Fig. \ref{fig:2b}, we visualize the local volume fraction $\phi$ in a region below the impactor (see Fig. \ref{fig:2a}).

The coarse-grained momentum density $\bm{p} (\bm{r},t)$ is written as
\begin{equation}
    \bm{p} (\bm{r},t) = \sum_i m_i \bm{u}_i (t) \Phi (\bm{r} - \bm{r}_i (t)),
    \label{eq:cg_p}
\end{equation}
where $\bm{u}_i$ and $m_i$ is the velocity and mass of particle $i$, respectively. 
The velocity field $\bm{u} (\bm{r},t)$ is defined by $\bm{u} (\bm{r},t) = \bm{p} (\bm{r},t) / \rho (\bm{r},t) $ with $\rho(\bm{r},t) := \sum_i m_i \Phi (\bm{r} - \bm{r}_i (t))$.
The stress tensor $\overleftrightarrow{\sigma}$ consists of the contact stress $\overleftrightarrow{\sigma}^c$ and the hydrodynamic stress $\overleftrightarrow{\sigma}^h$
\begin{equation}
    \overleftrightarrow{\sigma} :=  \overleftrightarrow{\sigma}^c + \overleftrightarrow{\sigma}^h.
\end{equation}
Here, the contact stress is expressed as 
\begin{equation}
    \overleftrightarrow{\sigma}^c (\bm{r}) = - \frac{1}{2} \sum_{i,j} \bm{F}^{c}_{ij} \otimes \bm{r}_{ij} \Phi(\bm{r}-\bm{r}_i)
\end{equation}
where $\bm{F}^{c}_{ij}$ and $\bm{r}_{ij}$ are the pairwise contact force and the interparticle distance between particles $i$ and $j$, respectively.
Here $\otimes$ denotes the tensor product.
Meanwhile, the hydrodynamic stress is given by
\begin{equation}
    \overleftrightarrow{\sigma}^h (\bm{r}) = \sum_i \overleftrightarrow{\sigma}_i^{h}\Phi(\bm{r}-\bm{r}_i),
\end{equation}
where $\overleftrightarrow{\sigma}_i^{h}$ is the hydrodynamic stress tensor on each particle, obtained from the LBM and the lubrication stresslet \cite{pradipto2021}.

The displacement field $\bm{U} (\bm{r})$ is defined by the particle displacement $\bm{U}_i (\bm{r}_i)$ from the equilibrium position as
\begin{equation}
    \bm{U} (\bm{r}) = \sum_i  \bm{U}_i \Phi(\bm{r}-\bm{r}_i) .
\end{equation}
Here, $\bm{U}_i$ is calculated as follows:
For each time $t$, an additional equilibration step is introduced where we freeze the motion of the impactor and allow the suspended particles to equilibrate.
Thus, the particle configuration from the LBM-DEM simulation at each time $t$, $\bm{r}_i (t)$, is considered as the initial condition in the equilibration process, i.e. $\bm{r}_i (t,s=0)$. 
During the equilibration process from $s$ to $s+\Delta s$, $\bm{r}_i (t,s)$ is updated considering only the hydrodynamic lubrication, normal and tangential contact forces until the equilibrium condition is reached where the average overlap between particles $\langle \delta_n^i \rangle$ is less than a threshold $\delta^{\rm th}$ at $s=s_{\rm th}$.
Then, the displacement from the equilibrium of particle $i$ at time $t$ is defined as
\begin{equation}
    \bm{U}_i (t) :=   \bm{r}_i (t;s=0)- \bm{r}_i (t; s=s_{\rm th}),
    \label{eq:dis}
\end{equation}
where the second term on the RHS of Eq. \eqref{eq:dis} is the equilibrated position.

Once the flow field is obtained, the symmetric part of the local strain rate tensor can be obtained $\overleftrightarrow{D} (\bm{r})$.
\begin{equation}
    \overleftrightarrow{D} (\bm{r}):= \frac{1}{2} ( \nabla \bm{u} + \nabla \bm{u}^{T} ).
\end{equation}
Meanwhile, the local strain tensor $\overleftrightarrow{L} (\bm{r})$ is defined as
\begin{equation}
\overleftrightarrow{L} (\bm{r}):= \frac{1}{2} ( \nabla \bm{U} + \nabla \bm{U}^{T} ).   
\end{equation}

Let us introduce the scalar local viscosity $ \eta (\bm{r}) $ defined as \cite{giusteri2018,giusteri2021},
\begin{equation}
    \eta(\bm{r}) := \frac{1}{2} \frac{\overleftrightarrow{\sigma}(\bm{r}):\overleftrightarrow{D}(\bm{r})}{\overleftrightarrow{D}(\bm{r}):\overleftrightarrow{D}(\bm{r})},
    \label{eq:eta}
\end{equation}
where $:$ is the scalar or double inner product.
The local strain rate $\dot{\varepsilon}$ is defined as
\begin{equation}
    \dot{\varepsilon} (\bm{r}) := \sqrt{2\overleftrightarrow{D}(\bm{r}):\overleftrightarrow{D}(\bm{r})}.
\end{equation}
A snapshot of the local strain rate field $\dot{\varepsilon} (\bm{r})$ right after an impact is shown in Fig. \ref{fig:2c}.
It can be seen that the position of the high rate region in our simulation is reminiscent of that observed experimentally in Ref. \cite{han2015}, although they used a constant penetrating intruder.
Then the local viscous stress $\sigma^{\rm (vis)}$ is simply given by \cite{giusteri2018,giusteri2021}
\begin{equation}
    \sigma^{\rm (vis)} (\bm{r}) = \eta (\bm{r}) \dot{\varepsilon} (\bm{r}).
\end{equation}
Finally, similar to the strain rate, the local scalar strain fields $\varepsilon(\bm{r})$ are defined as
\begin{equation}
    \varepsilon(\bm{r}) := \sqrt{2\overleftrightarrow{L}(\bm{r}):\overleftrightarrow{L}(\bm{r})}.
\end{equation}
A snapshot of the local strain field $\varepsilon(\bm{r})$ right after the impact is shown in Fig.\ref{fig:2d}.
Similar to Eq. \eqref{eq:eta} the local rigidity can be expressed as
\begin{equation}
G (\bm{r}):=\frac{1}{2}\frac{\overleftrightarrow{\sigma}(\bm{r}):\overleftrightarrow{L}(\bm{r})}{\overleftrightarrow{L}(\bm{r}):\overleftrightarrow{L}(\bm{r})}.
\end{equation}
The local elastic stress is then given by
\begin{equation}
\sigma^{(\rm el)}(\bm{r})=G(\bm{r})\varepsilon(\bm{r}).
\end{equation}
In Fig. \ref{fig:2e}, we visualize the total scalar stress $\sigma = \sigma^{(\rm el)} + \sigma^{(\rm vis)}$ after the impact.


Once we have computed the local variables within the suspensions, delineating the submerged impactor surface $S$ with normals $\bm{n}$ (see Appendix \ref{app:surf} for details), one can evaluate the effective volume fraction around the impactor $\phi_{\rm eff}$ defined as
\begin{equation}
    \phi_{\rm eff}:= \frac{\int_S \phi (\bm{r}) dS}{\int_S dS}.
    \label{eq:phi_eff}
\end{equation}
where $dS$ is the surface integration on $S$.
Similarly, the effective viscosity around the impactor $\eta_{\rm eff}$ is defined as
\begin{equation}
        \eta_{\rm eff}:= \frac{\int_S \eta (\bm{r}) dS}{\int_S dS} .
        \label{eq:eta_eff}
\end{equation}  
The time evolution of $\phi_{\rm eff}$ can be seen in Fig. \ref{fig:3b}, where $\phi_{\rm eff}$ increases right after the impact.
We have also plotted the time evolution of $\eta_{\rm eff}$ in Fig. \ref{fig:3c}, where it also increases right after the impact.
This suggests that the effective viscosity $\eta_{\rm eff}$ satisfies the constitutive law for viscosity \cite{boyer2011,suzuki2019}
\begin{equation}
   \frac{\eta_{\rm eff}  
}{\eta_0 }=\bigg( \frac{\phi_{\rm eff} (\phi_J - \phi_{0})}{\phi_{0}(\phi_J - \phi_{\rm eff} )} \bigg)^2,
\label{eq:eta_eff_m}
\end{equation}
where $\phi_J$ is the volume fraction at the jamming point, $\phi_{\rm eff}$ is the effective volume fraction around the impactor, and $\phi_0$ is the initial volume fraction.
Note that $\phi_\mathrm{eff}$ is larger than $\phi_0$ because the impactor makes DJR right below it.
In Fig. \ref{fig:3c}, we also plot Eq. \eqref{eq:eta_eff_m} as dashed lines, where the measurement (Eq. \eqref{eq:eta_eff}) agrees with the empirical expression (Eq. \eqref{eq:eta_eff_m}).

\subsection{Force acting on the impactor}\label{SecIID}

\begin{figure*}[htbp]
    \centering
	\subfloat[]{\label{fig:4a}%
		\includegraphics[width=0.4\linewidth]{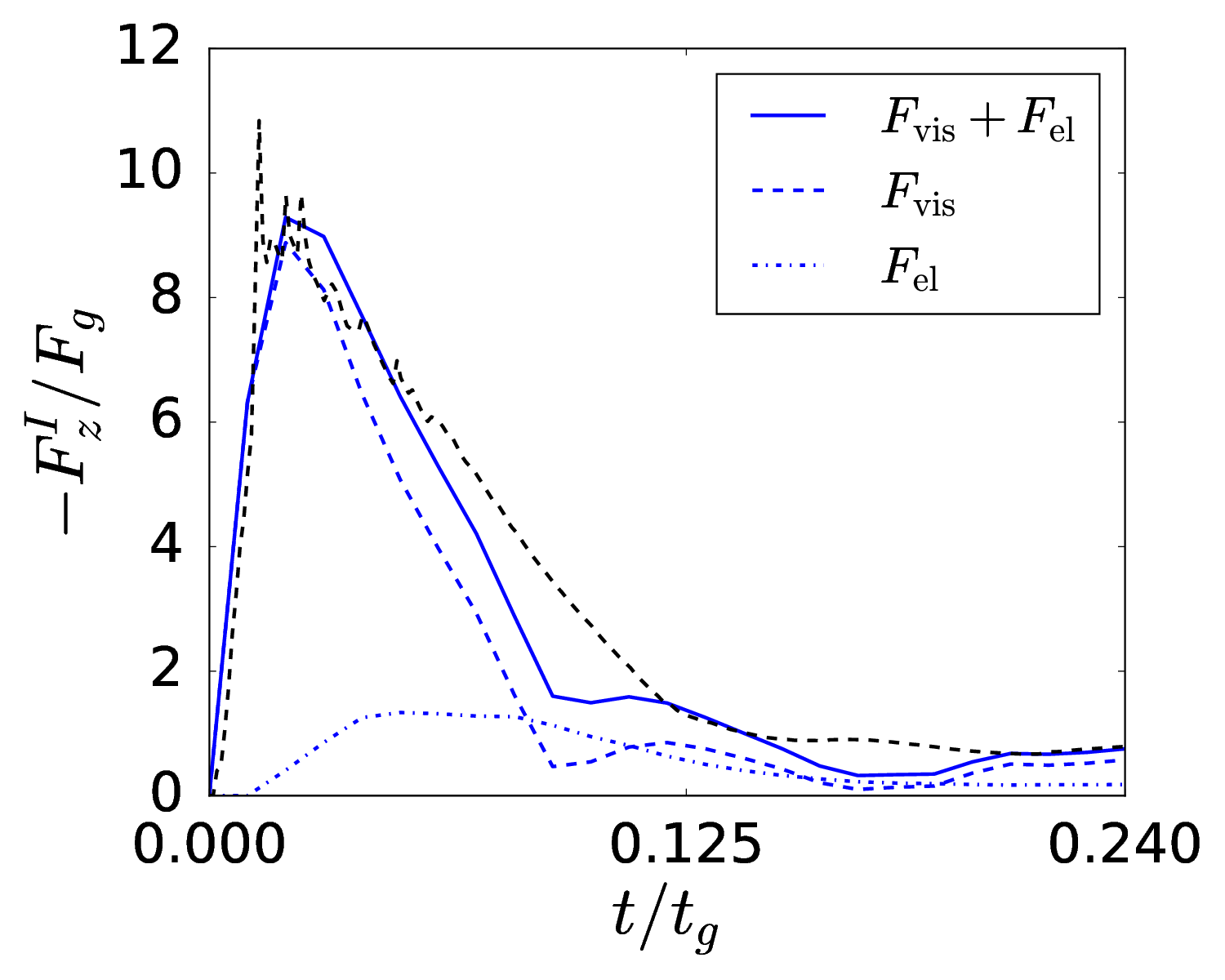}
	}
		\subfloat[]{\label{fig:4b}%
		\includegraphics[width=0.4\linewidth]{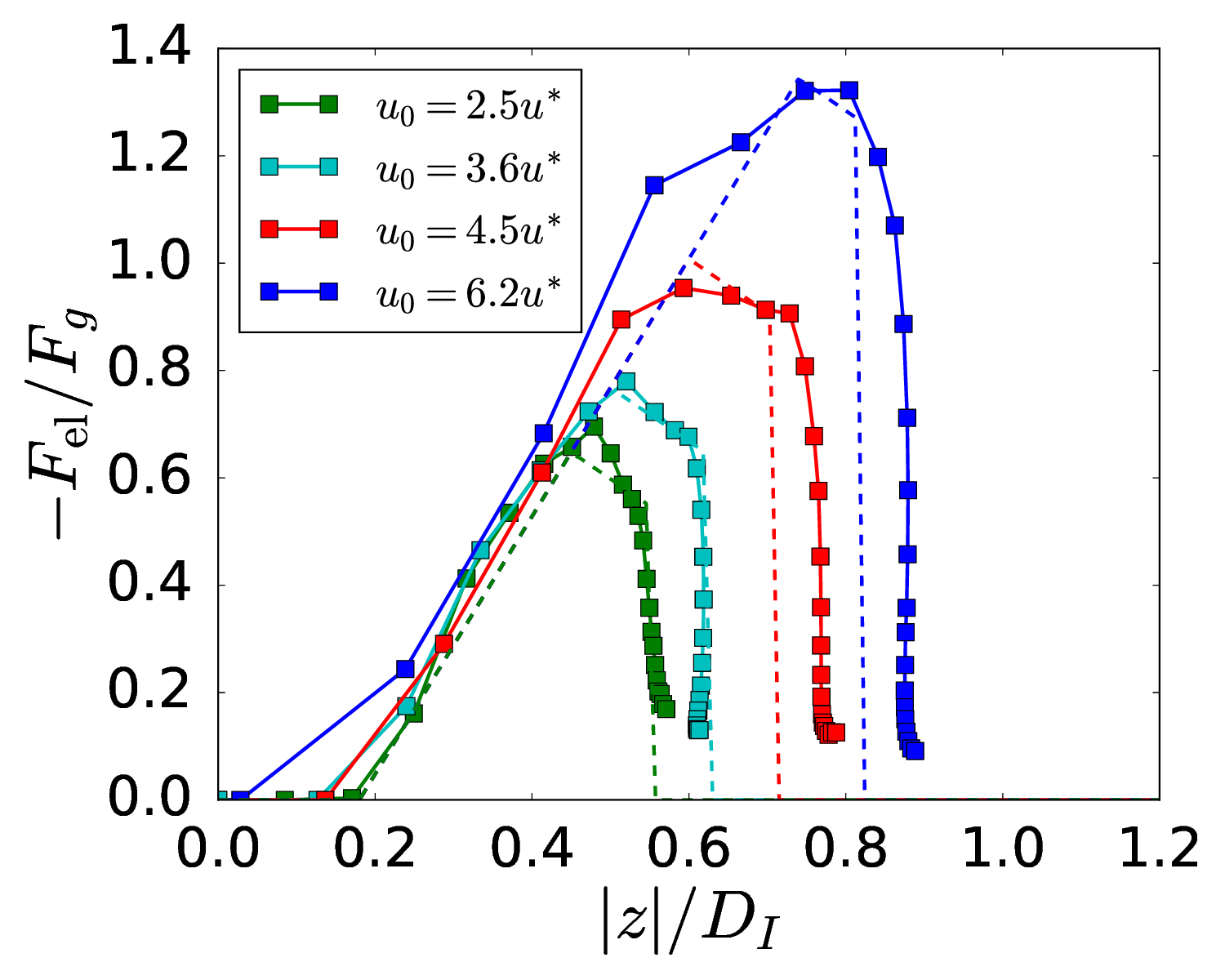}
	}
	\caption{(a)Plots of the time evolution of the force acting on the impactor by LBM-DEM (black-dashed line), the viscous force expressed as Eq. \eqref{eq:f_vis} (blue dashed line), the elastic force expressed as Eq. \eqref{eq:f_el} (blue dotted line), and their summations (blue solid line).
 (b) Plots of elastic force against depth for various impact velocities, where the dashed lines express Eq. \eqref{eq:fel} with numerically evaluated $z_{\rm me}$ and $z_{\rm cut}$ by setting $z_{\rm on} = -0.18D_I$.}
	\label{fig:4}
\end{figure*}

The force acting on the impactor can be obtained by integrating the stress field on the surface $S$, 
where the viscous force and elastic force are, respectively, given by 
\begin{align}
    F_{\rm vis} &=  \int_S \sigma^{(\rm vis)} \bm{n} dS,  \label{eq:f_vis}\\
    F_{\rm el} &=  \int_S \sigma^{(\rm el)} \bm{n} dS,  \label{eq:f_el}
\end{align}
In Fig. \ref{fig:4a}, we plot the total force obtained by Eqs.~\eqref{eq:f_vis} and \eqref{eq:f_el} (blue solid line) and compare it with the force measured directly on the impactor using LBM-DEM simulation (black dashed line).
Although the coarse-grained method cannot be used for sharp impulses in a short time, the agreement between the two methods is reasonable.
As reported in Ref.~\cite{pradipto2021b}, the viscous force dominates, in particular, in the early stage, but the elastic force plays an important role after the time to take the peak of the force.
Although percolating force chains of suspended particles do not exist for this parameter setup (see the time evolution of the force chains in the Supplemental Movie \cite{supp_fc_u4}), it is noteworthy that the elastic force still exists.

To get a better understanding of the elastic force, we plot $F_{\rm el}$ against the normalized depth $|z|/D_I$ for various $u_0$ with $\phi_0 = 0.53$ in Fig. \ref{fig:4b}.
As can be seen, the onset depth $z_\mathrm{on}$ of the elastic force $F_\mathrm{el}$ little depends on $u_0$.
Then $F_\mathrm{el}$ increases linearly with $|z|$ until reaching the maximum value at certain $z_{\rm me}$ which depends on $u_0$.
For $|z|>|z_\mathrm{me}|$, $F_\mathrm{el}$ likely decreases almost linearly with $|z|$ for $|z|<|z_\mathrm{cut}|$, at least, for small $u_0$, and $F_\mathrm{el}$ suddenly drops to zero.

Based on these observations, we propose the following empirical expression for the elastic force $F_{\rm el}$ 
\begin{widetext}
\begin{equation}
F_{\rm el} (z) = \begin{cases}
0, \qquad  |z| < |z_{\rm on}|, \\
k(z - z_{\rm on}), \qquad | z_{\rm on}| \leq |z| \leq | z_{\rm me}|,\\
 -k^{\prime} (z - z_{\rm me}) +k(z_{\rm me}- z_{\rm on}), \qquad  |z_{\rm me}| < |z| \leq |z_{\rm cut}|, \\
0, \qquad |z| > |z_{\rm cut}|,
\end{cases}
\label{eq:fel}
\end{equation}
\end{widetext}
where $z_{\rm on}$, $z_{\rm me}$, $z_{\rm cut}$ are the position of the onset of elastic force, the position of maximum elastic force, and the cut-off position of the elastic force, respectively.
Here, $k$ and $k^{\prime}$ are fitting parameters that are related to the stiffness of the suspended particles. 
We treat $z_{\rm on}$ as a fitting parameter and based on Fig. \ref{fig:4b}, we choose $z_{\rm on} = -0.18 D_I$.
Then, $k$ can be estimated by fitting the data with a linear function i. e. the second equation of Eq. \eqref{eq:fel}, where we estimate $k=260 m_0 / (a_{\rm min} t_{g}^{2})$.
Similarly, with the third equation of Eq. \eqref{eq:fel}, we estimate $k^{\prime} = 110 m_0 / (a_{\rm min} t_{g}^{2})$.

\subsection{Reduced equation of motion}\label{SecIIE}

\begin{figure*}[htbp]
    \centering
	\subfloat[]{\label{fig:5a}%
		\includegraphics[width=0.4\linewidth]{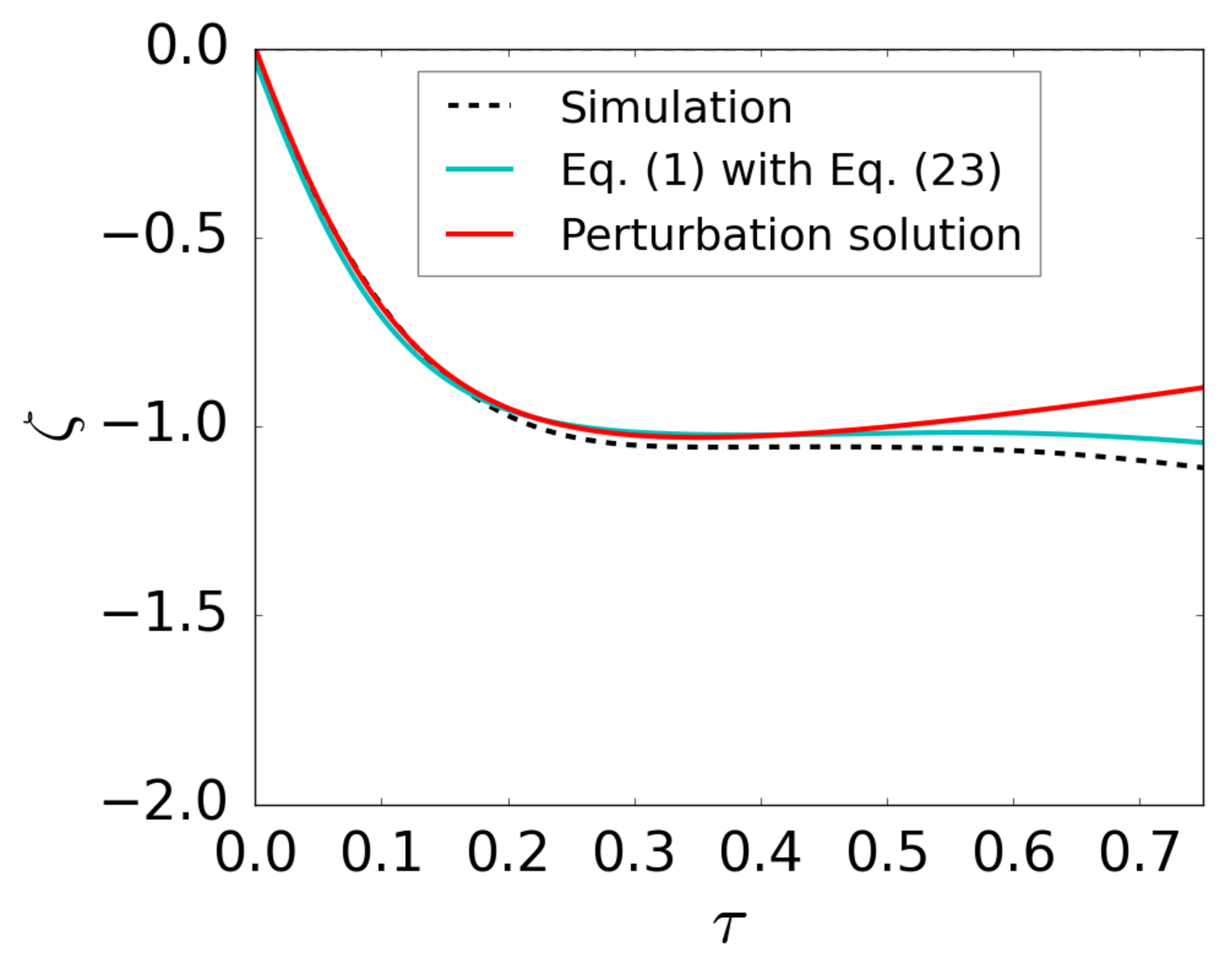}
	}
		\subfloat[]{\label{fig:5b}%
		\includegraphics[width=0.4\linewidth]{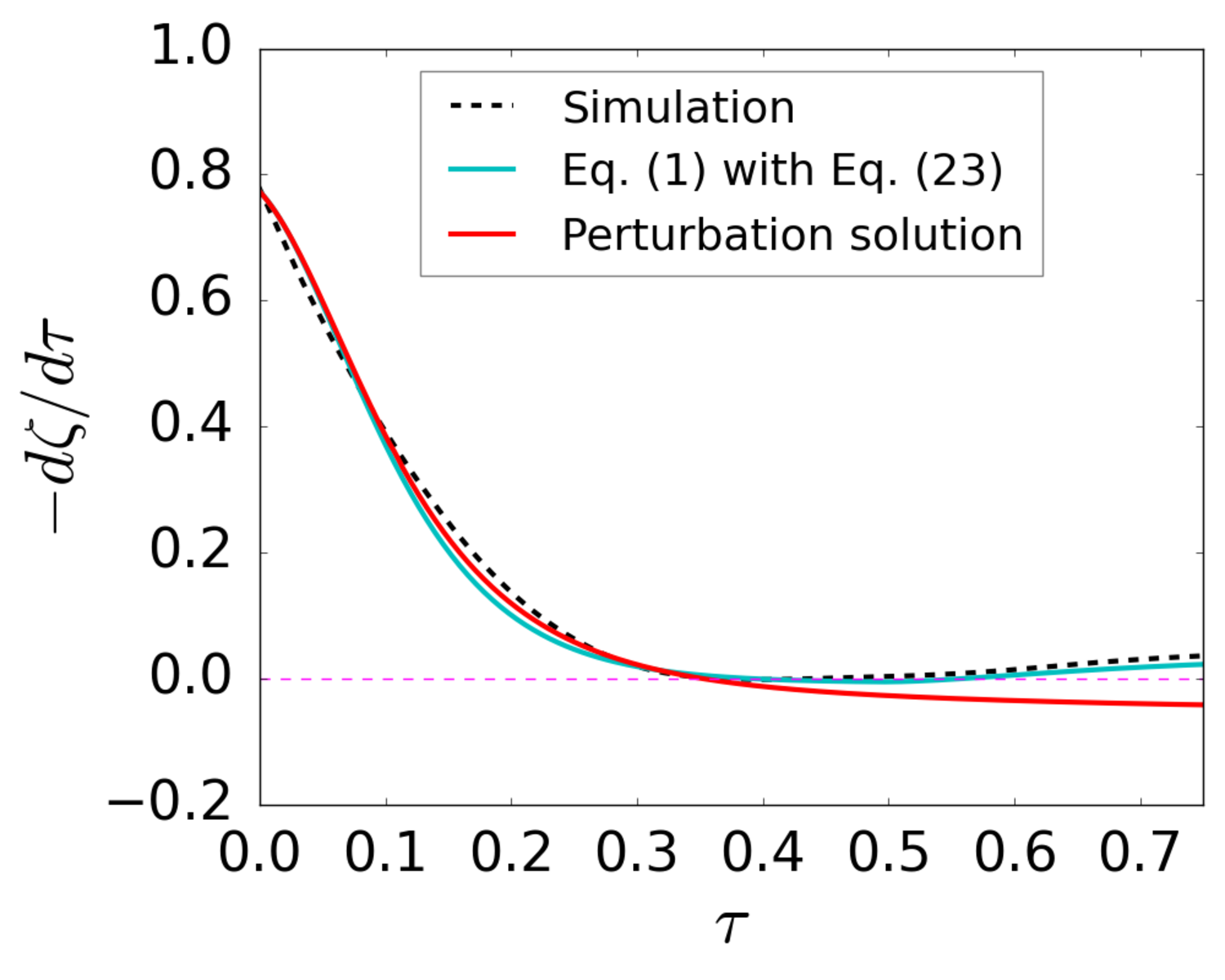}
	}
	\caption{Plots of the time evolutions of $\zeta=z/a_I$ (a) and $d\zeta/d\tau$ with  $\tau:=3\pi \eta_{\rm eff}a_I t/m_I$ (b).
Here, the red solid lines, green solid lines and dotted lines correspond to
 the solution of the perturbative equation, the solution of the reduced model (Eqs. \eqref{eq:eom} and \eqref{eq:fel} with $z_{\rm on} = -0.18D_I$ and numerically evaluated $z_{\rm me}$ and $z_{\rm cut}$), and the LBM-DEM simulation results, respectively.
 }
	\label{fig:5}
\end{figure*}

The next task is to determine $z_{\rm me}$ and $z_{\rm cut}$.
Using the second expression for $F_\mathrm{el}(z)$, Eq. \eqref{eq:eom} can be rewritten as
\begin{equation}
    m_I \ddot{z} = - m_I \tilde{g} - 3 \pi \eta_{\rm eff} \dot{z} |z| - k(z - z_{\rm on}) \qquad t_{\rm on} \leq t < t_{\rm me}.
    \label{eq:eom2}
\end{equation}
It is obvious that the maximum elastic force occurs when the sign of impactor velocity switches ($\dot{z}=0$).
Because the full solution of Eq. \eqref{eq:eom2} is complicated, it is impossible to determine $z_{\rm me}$ ($z$-position corresponding to $\dot{z}=0$) analytically.
Thus, we solve Eq. \eqref{eq:eom2} numerically and obtain $t_{\rm me}$ with $z_{\rm me} =  z(t_{\rm me})$ as the time satisfying $\dot{z} = 0$.
Note that the elastic force is continuous at $z_{\rm me}$.

For $|z_{\rm me}| < |z| < |z_{\rm cut}|$, the equation of motion can be written as
\begin{equation}
        m_I \ddot{z} = - m_I \tilde{g} - 3 \pi \eta_{\rm eff} \dot{z} |z| - k (z_{\rm me} - z_{\rm on}) + k^{\prime}(z-z_{\rm me}).
        \label{eq:eom3}
\end{equation}
The mechanical energy of the system $E$, consisting of kinetic and elastic energy is given by
\begin{equation}
   E = \frac{1}{2} m_I \dot{z}^2 + \frac{1}{2} k^{\prime} (z - z_{\rm me})^2.
    \label{eq:e}
\end{equation}
Multiply Eq. \eqref{eq:eom3} with $\dot{z}$, one gets
\begin{align}
    \frac{dE}{dt} &= - \dot{z} (m_I g + k (z_{\rm me}-z_{\rm on})) - 3 \pi \eta \dot{z}^2 |z| \notag\\ &+ 2 k^{\prime} \dot{z} (z - z_{\rm me}) .
    \label{eq:ener}
\end{align}
The restoring potential energy $E_{\rm me}: = E (t=t_{\rm me}) = \frac{1}{2} k z_{\rm me}^2$ is completely dissipated at $t_{\rm cut}$ and $z_\mathrm{cut}:=z(t=t_\mathrm{cut})$.
Thus, $z_{\rm cut}$ and $t_\mathrm{cut}$ are determined by
\begin{widetext}
\begin{align}
\frac{1}{2} k z_{\rm me}^2   = \int_{t=t_{\rm me}}^{t=t_{\rm cut}}  \{ \dot{z}(t) (m_I g + k (z_{\rm me}-z_{\rm on}))  + 3 \pi \eta \dot{z}^2(t) |z(t)|  - 2 k^{\prime} \dot{z}(t) (z(t) - z_{\rm me})\}dt . 
\label{eq:tcut}
\end{align}
\end{widetext}
Figure \ref{fig:4b} displays both the empirical expression Eq.~\eqref{eq:fel} for $F_\mathrm{el}$ acting on the impactor and that by LBM-DEM with the aid of numerically evaluated $z_{\rm me}$ and $z_{\rm cut}$ with fitting parameters $z_{\rm on}$, $k$, and $k^{\prime}$.
This indicates that our empirical expression is a reasonable one for the elastic force $F_\mathrm{el}$.

Once we estimate all parameters in Eq. \eqref{eq:fel} and the effective viscosity as in Eq. \eqref{eq:eta_eff_m}, one can solve Eq. \eqref{eq:eom} numerically with the Adams-Bashforth method with the time increment $\Delta t =10^{-3} t_g$ and the local error $r_{\rm tol} = 10^{-8}$ \cite{lsoda}.
As can be seen in Fig. \ref{fig:5}, we get a good agreement between the solution of the full LBM-DEM simulation and the solution of Eq. \eqref{eq:eom} with Eq. \eqref{eq:fel}.
We also compare the perturbation solution of Eq. \eqref{eq:eom2} in which the elastic force is treated as a perturbation to the dominant viscous force (details in Appendix \ref{app:perturbation}) with the results obtained by the other methods.
It seems the perturbation works well for $\tau :=  3\pi \eta_{\rm eff}a_I t/m_I < 0.4$.
After the impactor reaches the minimum velocity, it starts to sink due to the relaxation of the suspensions.
Indeed, the sinking/relaxation process is currently ignored in our perturbation approach.

Our results help us to understand the origin of elasticity in suspensions without percolating clusters of contacted particles.
From the method to evaluate the elastic force, the restoring force from the displacement of the suspended particles, Eq. (\ref{eq:dis}), induced by an impact process to a stable configuration can be regarded as the elastic force.
Indeed, the suspended grains under a finite speed impact are moved in unstable configurations, and they are relaxed to the stable configuration as time goes on. 
In other words, the elasticity in dense suspensions disappears in quasi-static processes.

\section{Foot-spring-body dynamics in dense suspensions}\label{SecIII}

\begin{figure*}[htbp]
    \centering
	\subfloat[]{\label{fig:6a}%
		\includegraphics[width=0.3\linewidth]{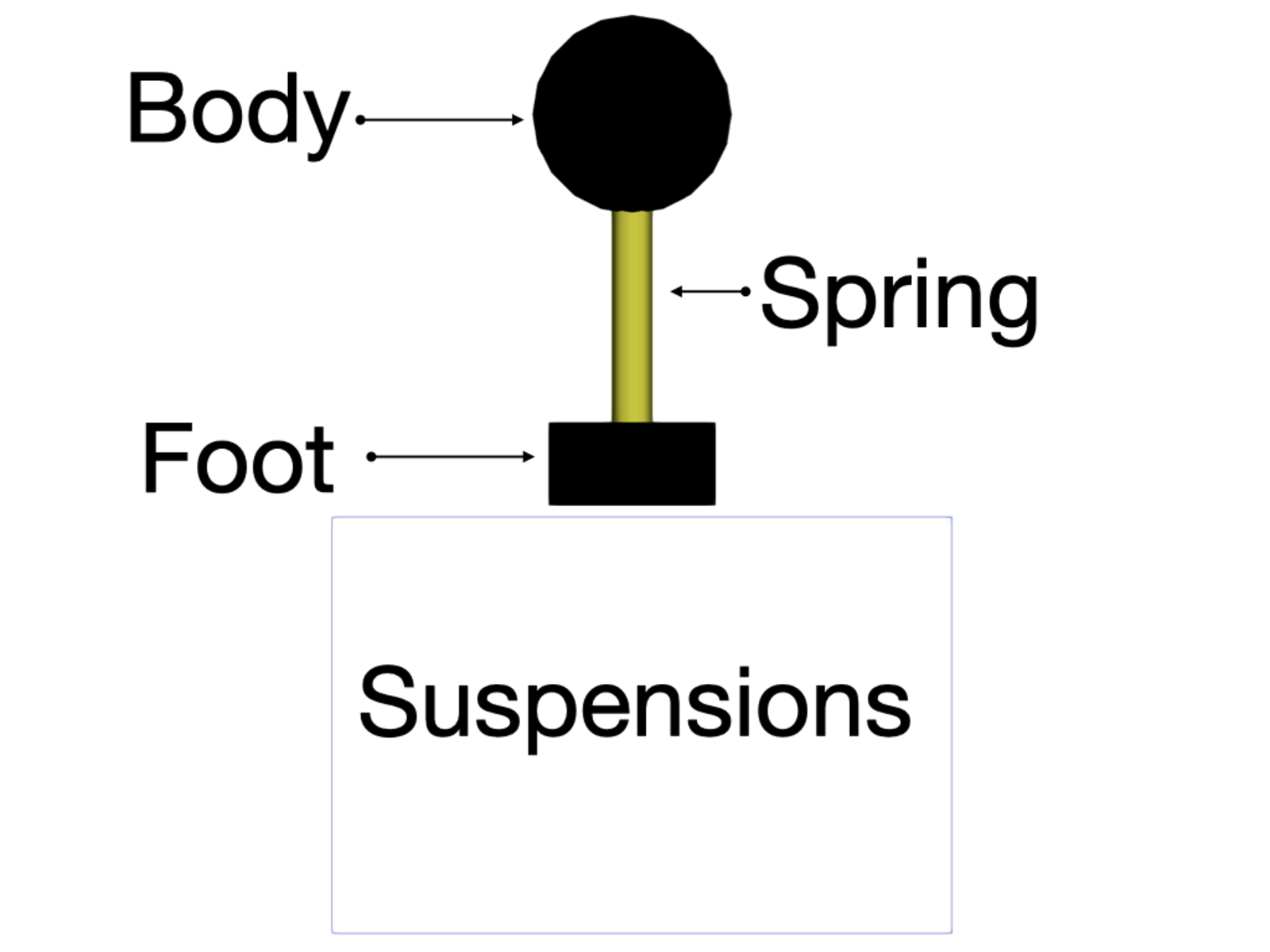}
	}
	\subfloat[]{\label{fig:6b}%
		\includegraphics[width=0.3\linewidth]{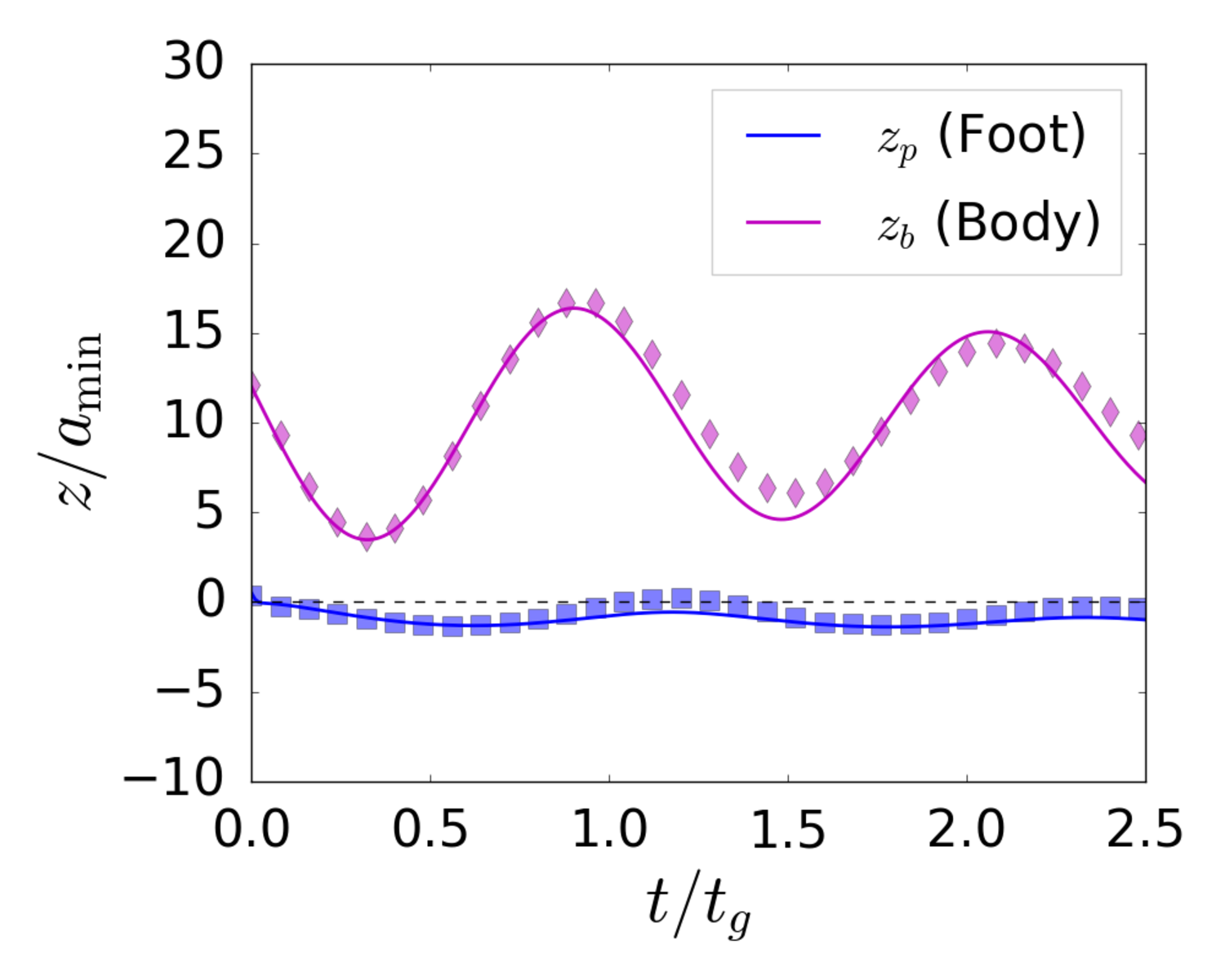}
	}
 	\subfloat[]{\label{fig:6c}%
		\includegraphics[width=0.3\linewidth]{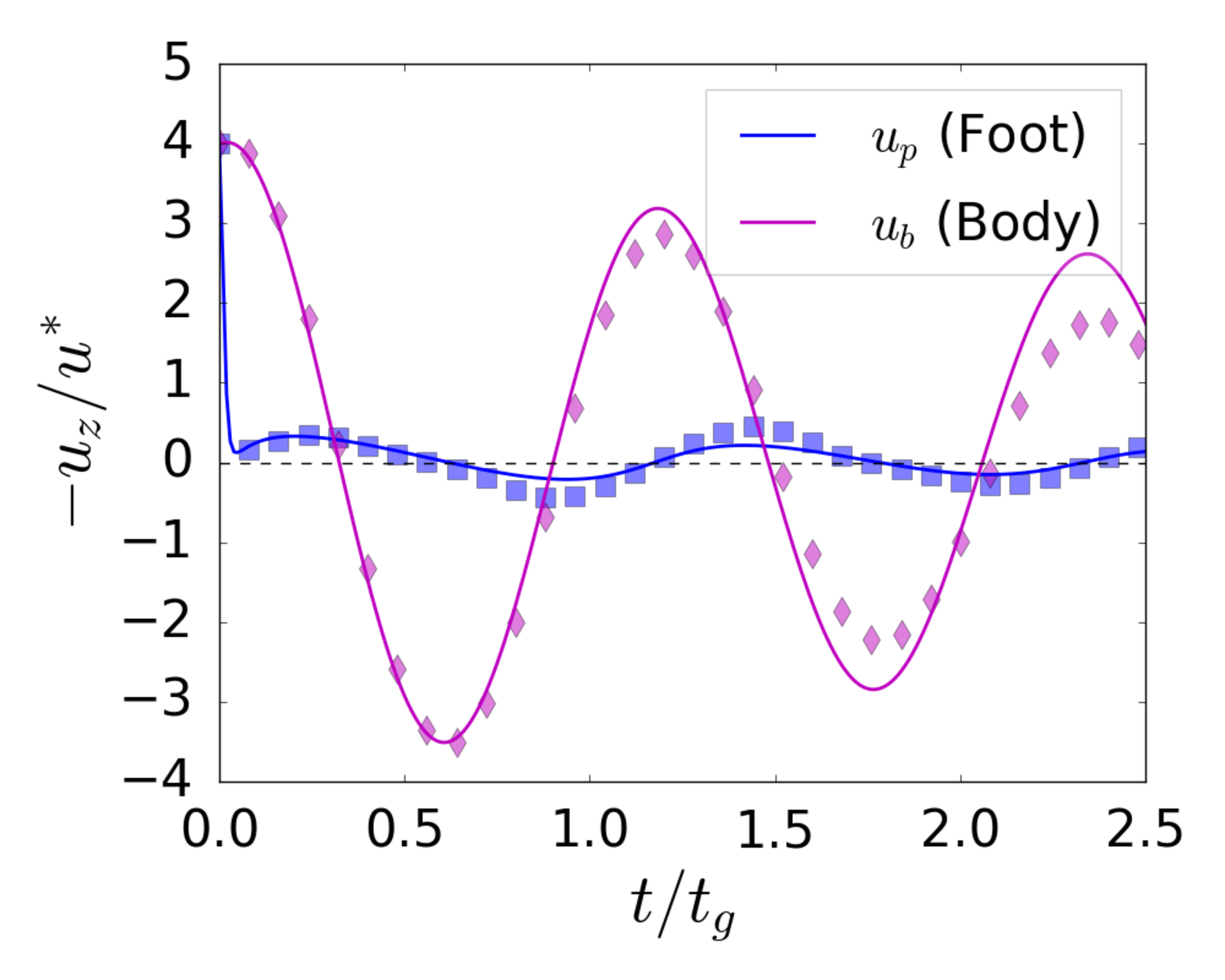}
	}
 
 	\subfloat[]{\label{fig:6d}%
		\includegraphics[width=0.4\linewidth]{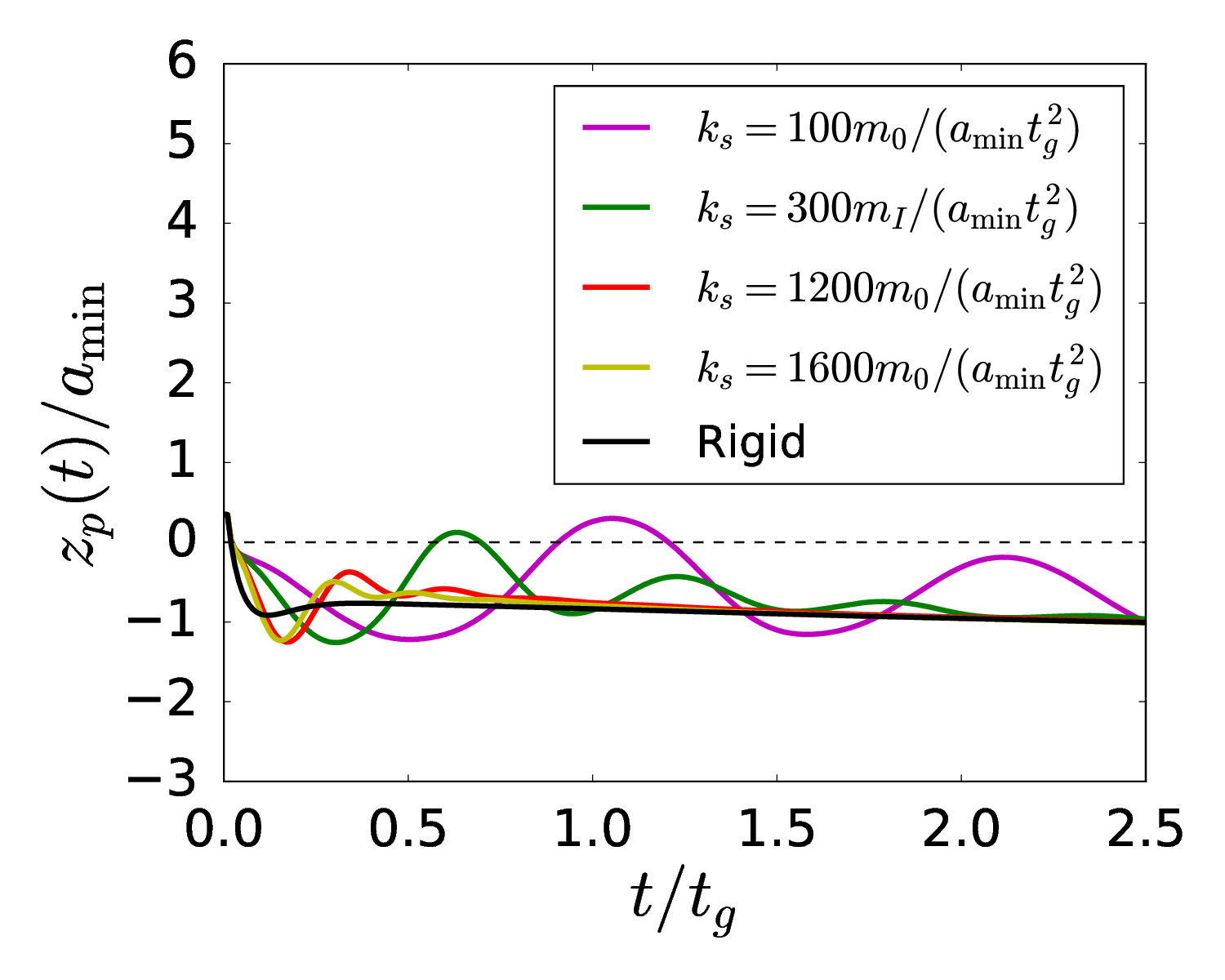}
	}
 	\subfloat[]{\label{fig:6e}%
		\includegraphics[width=0.4\linewidth]{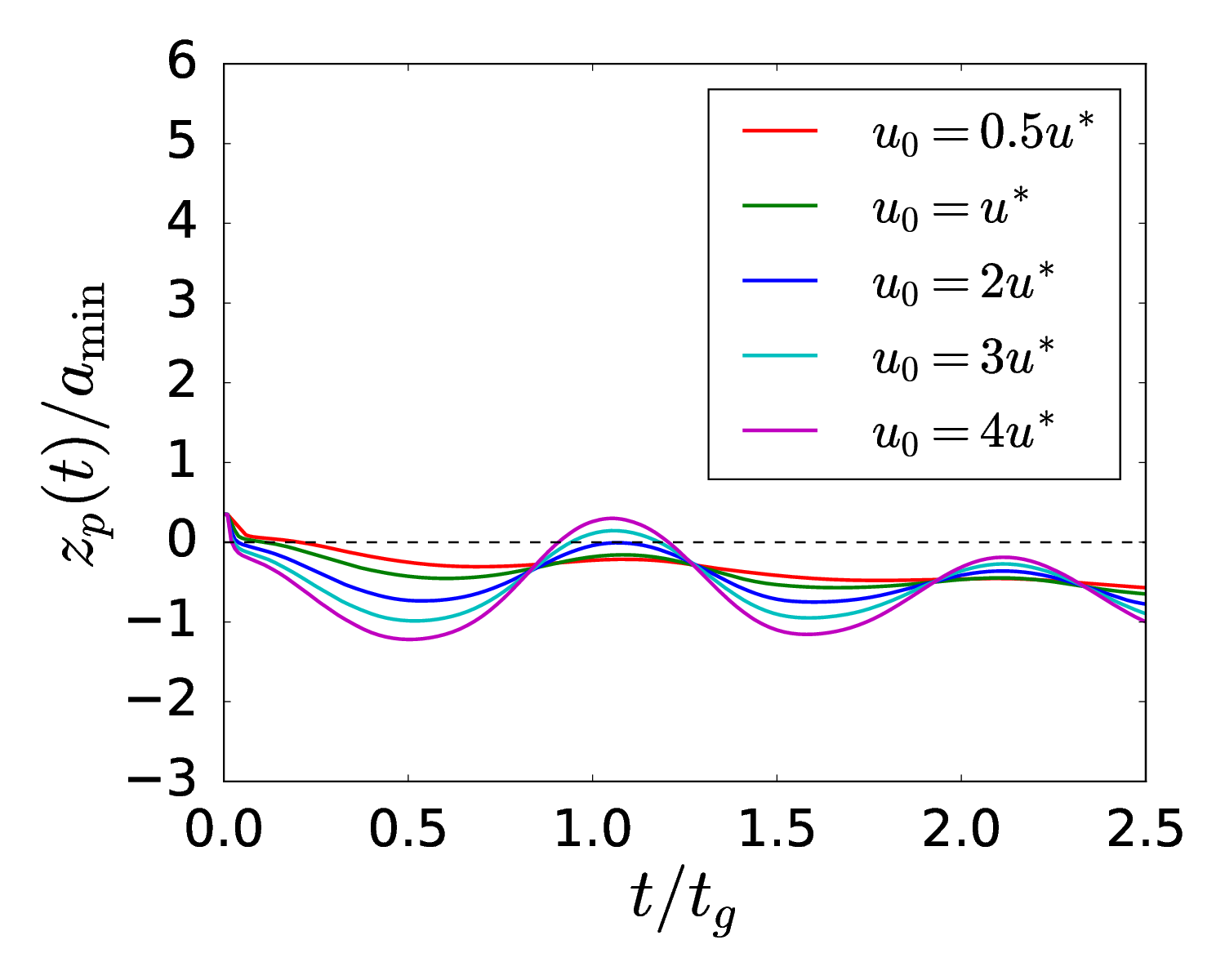}
	}
	\caption{ (a) A schematic of the simulation setup of the foot-spring-body system. The body (black circle) is connected with the foot (black rectangle) by a massless spring (yellow tube). 
 (b) Time evolution of the velocities of the foot (blues) and body (purples) in $z-$coordinate, where the black dashed line expresses $u_z= 0$, the solid lines express the solutions of Eq. \eqref{eq:eom_spring}, and the symbols express the results of the simulations. 
 (c) Time evolution of the positions of the foot (blues) and body (purples) in $z-$coordinate, where the black dashed line expresses the suspension surface, the solid lines express the solutions of Eq. \eqref{eq:eom_spring}, and the symbols express the results of the LBM-DEM simulations. All results here are obtained with $k_s=100 m_0 / (a_{\rm min} t_g^2 )$ and $u_0 =4 u^{*}$.
 (d) Time evolution of the foot position in $z-$direction $z_p$ with various spring stiffness $k_s$ at $u_0 = 4u^{*}$. 
 (e) Time evolution of the foot position in $z-$direction $z_p$ with various initial velocity $u_0$ at $k_s =100 m_0 / (a_{\rm min} t_g^2)$.
 }
	\label{fig:6}
\end{figure*}

The motivation of our study is to mimic walking processes on dense suspensions using a simple model.
For this purpose, we introduce the foot-spring-body model as a model for expressing the bouncing motion on a suspended liquid.
In this section, we explain the model to examine whether the model can reproduce multiple bounces after dropping it on the suspensions based on the LBM-DEM simulation.
We also adopt a reduced model as in the previous section and demonstrate that the model can mimic walking on the suspension.

The foot in the foot-spring-body model is represented by a rectangular plate impactor with volume $V_p := W_p \times H_p \times D_p$ and mass $m_p =\rho_p  V_p $, where $\rho_p$ is the density of the footplate.
We adopt $\rho_p = 1.2 \rho_f$ and $W_p=D_p=5a_{\rm min}$ and $H_p= 2a_{\rm min}$.
The body is represented by a sphere with diameter $D_b$ and mass $m_b$.
We take the density of the body $\rho_b$ as $\rho_b=2 \rho_p =2.4\rho_f$.
The body and the foot are then connected by a massless spring with stiffness $k_s$ and natural length $L_0$.
The schematic of this setup is shown in figure \ref{fig:6a}.
Even for a high-volume fraction, the impactor sinks eventually over a long time limit.
Such sinking can be avoided if we introduce an internal degree of freedom on the impactor, such as a spring introduced here.
In order to reduce the simulation time for sinking processes, we adopt a slightly lower volume fraction $\phi_0 = 0.51$ than that in the previous section for the analysis.
Here we use $H=2D_b$, $W=D=4D_b$ and $N=618$.
Note that we are only interested in the vertical ($z$ direction) motion of the system.

Thus, a reduced set of equations for the foot-spring-body model corresponding to Eq. \eqref{eq:eom} is given by
\begin{align}
        m_b \ddot{z}_b = & -m_b g - k_s (z_b - z_p - L_0) - \zeta_s \dot{z}_b \notag\\
        m_p \ddot{z}_p  = & -m_p g - 3 \pi \eta_{\rm eff} z_p \dot{z}_p + F_{\rm el} (z_p) \notag\\ &+ k_s (z_b - z_p - L_0)  - \zeta_s \dot{z}_p,
    \label{eq:eom_spring}
\end{align}
where $m_b$ and $z_b$ are the mass and the vertical position of the body, respectively.  
$z_p$ is the vertical position of the base of the plate impactor, and $\zeta_s:=\sqrt{k_s (m_p + m_b)/2}$ is the damping constant.
Typical motions of the foot-spring-body system are shown in Figs. \ref{fig:6b} and \ref{fig:6c}.
To solve Eq. \eqref{eq:eom_spring}, $\eta_{\rm eff}$ is estimated for the short time using Eq. \eqref{eq:eta_eff_m} as in the previous section, while Eq. \eqref{eq:fel} with a new set of parameters is used to recover the elastic force $F_{\rm el} (z_p)$.
Here, we adopt $z_{\rm on}= 0.25 W_p$, $k=214 m_0 / (a_{\rm min} t_{g}^{2})$, and $k^{\prime}=114 m_0 / (a_{\rm min} t_{g}^{2})$.
Then, $z_{\rm me}$ and $z_{\rm cut}$ are obtained using a parallel procedure in the previous section.
We then solve Eq. \eqref{eq:eom_spring} numerically.
As can be seen, the solution of Eq. \eqref{eq:eom_spring} agrees well with the simulation results.
Thus, our reduced model Eq. \eqref{eq:eom_spring} is a reasonable model to analyze the motion of the foot-spring-body model.
Initially, the foot experiences a strong deceleration as in the free-falling impactor due to the interaction between the foot and the suspensions.
Meanwhile, the body continues to accelerate due to gravity.
Then, the system exhibits a damped oscillation.
Due to the spring force and the stiffness of the suspensions, the foot undergoes multiple bounces ($u_p <0$) and also multiple jumps ($z_p > 0$).
This result suggests that composites with elastic springs inside the body can maintain their position above the liquid surface for a while.

Now, let us investigate the multiple bounces of the foot in detail.
First, we check how the motion of the foot depends on the stiffness of the spring $k_s$.
The simulation results for various $k_s$ are shown in Fig. \ref{fig:6d}.
Here, one can see a lower tendency to multiple bounces for higher $k_s$.
Furthermore, the foot only bounces once and then sinks in a rigid limit ($k_s \rightarrow \infty$).
This is similar to the prediction of the added mass model in Ref. \cite{mukhopadhyay2018}, where running on suspensions is impossible for a perfectly stiff leg.
We also examine the dependence of the initial velocity ($u_0$) in Fig. \ref{fig:6e}.
As expected, the foot sinks and does not hop at low $u_0$, since the impact-induced hardening is stronger at high $u_0$ \cite{waitukaitis2012,egawa2019,brassard2020,pradipto2021}.

\section{Conclusions and discussions}\label{Conclusion}

Using the coarse-grained method and the virtual deformation of the suspended particles from unstable to equilibrium positions, we evaluate the  viscous and elastic forces acting on the impactor.
We found increases in viscosity and density around the impactor right after impact.
We confirmed that the elastic force acting on the impactor $F_{\rm el}$ exists even in the absence of percolating clusters of suspended particles.
The behavior of $F_{\rm el}$, which depends on depth $z$, can be expressed as an empirical equation with five fitting parameters (onset of elastic force $z_{\rm on}$, position of maximum elastic force $z_{\rm me}$, position $z_{\rm cut}$ where the initial mechanical energy $E$ becomes completely dissipated, spring constants $k$ and $k^{\prime}$).
Using this $F_\mathrm{el}$ with $F_\mathrm{vis}=- 3 \pi \eta_{\rm eff} \dot{z} |z|$ we obtain the reduced equation.
The solution of the reduced equation is almost equivalent to that for the full set of equations of LBM-DEM.
 
Finally, to mimic walking on a liquid, we studied the impact of the foot-spring-body system on the top of dense suspensions.
Our reduced model for this system agrees well with the results of the LBM-DEM simulation.
We confirmed that multiple bounces are suppressed as the spring stiffness $k_s$ between the body and foot increases and the initial velocity $u_0$ decreases.

We expect that our method is applicable to the sinking process of an intruder in dense suspensions, where oscillations and slip-stick motions have been observed \cite{vonkann2011}.
However, such a sinking process is beyond the scope of this paper.
Relatedly, little is known about the relaxation process of hardening suspensions after impact \cite{maharjan2017,cho2022,barik2022}.
Future studies should focus on this relaxation phenomenon of dense suspensions under impact.

\section*{Acknowledgements}
One of the authors (P.) expresses his gratitude to Alessandro Leonardi for sharing his lattice Boltzmann code.
We thank Ryohei Seto and Satoshi Takada for their useful comments.
One of the authors (P.) also thanks Yoshiyuki Tagawa for his support during the revision process of this paper.
This work is partially supported by the  Grant-in-Aid of MEXT for Scientific Research KAKENHI (Grant No. JP21H01006).
All numerical calculations were carried out at the Yukawa Institute for Theoretical Physics (YITP) Computer Facilities, Kyoto University, Japan.

\appendix
\section{\label{app:lbm} LBM-DEM with free surface}

We employ the LBM involving suspensions and the free surface of the fluid.
The details of the LBM are explained in Ref.~\cite{pradipto2021}.
The suspended particles in LBM are represented as a group of solid nodes, while the surrounding fluids are represented by fluid nodes.
The hydrodynamic field is calculated from the time evolution of the discrete distribution function at each fluid node.
We select the lattice unit $\Delta x_{\rm lb} = 0.2 a_{\text{min}}$, where it gives sufficient accuracy but is still not computationally expensive as shown in the previous LBM for suspensions literature \cite{ladd1994a,ladd1994b,nguyen2002}.
In addition, to simulate the free surface of the fluid, it is necessary to introduce interface nodes between the fluid and gas nodes \cite{svec2012,leonardi2014,leonardi2015, pradipto2021}.

Equations of motion and the torque balance of particle $i$ are, respectively, given by 
\begin{equation}
	m_i \frac{d \bm{u}_i}{dt} = \bm{F}_i^c + \bm{F}_i^h + \bm{F}_i^{\rm lub} + \bm{F}_i^r 
	\label{eom_part}
\end{equation}
\begin{equation}
	I_i \frac{d \bm{\omega}_i}{dt} = \bm{T}_i^c +  \bm{T}_i^{\rm lub} + \bm{T}_i^{h}.
	\label{eot_part}
\end{equation}
Here, $\bm{u}_i$, $\bm{\omega}_i$, $m_i$, and $I_i=(2/5) m_i a_i^2$ (with $a_i$ the radius of particle $i$), are the translational velocity, angular velocity, mass, and the moment of inertia of particle $i$, respectively.

Note that our LBM accounts for both the short-range lubrication force $\bm{F}_i^{\rm lub}$ and torque $\bm{T}_i^{\rm lub}$, as well as the long-range hydrodynamic force $\bm{F}_i^h$ and torque $\bm{T}_i^h$ as in Ref.~\cite{nguyen2002,pradipto2020}.
The long-range parts ($\bm{F}_i^h$ and $\bm{T}_i^h$) are calculated using the direct forcing method~\cite{leonardi2015,pradipto2021}, while the lubrication force $\bm{F}_i^{\rm lub}$ and torque $\bm{T}_i^{\rm lub}$ are expressed by pairwise interactions as $\bm{F}_{i}^{\rm lub} = \sum_{j \neq i} \bm{F}_{ij}^{\rm lub}$ and $\bm{T}_{i}^{\rm c} = \sum_{j \neq i} \bm{T}_{ij}^{\rm lub}$, respectively \cite{seto2013,mari2014,nguyen2002,pradipto2020}.
The explicit expressions of $\bm{F}_{ij}^{\rm lub}$ and $\bm{T}_{ij}^{\rm lub}$ can be found in Ref. \cite{pradipto2020}.

We adopt the linear spring-dashpot version of the DEM \cite{luding2008} for the contact interaction between particles, which involves both the normal and the tangential contact forces.
Note that we omit the dissipative part for the tangential contact force.
For the particle $i$, the contact force $\bm{F}^{c}_{i}$ and torque $\bm{T}^{c}_{i}$ are, respectively, written as $\bm{F}^{c}_{i} = \sum_{i \neq j} (\bm{F}^{\text{nor}}_{ij} + \bm{F}^{\text{tan}}_{ij} )$ and  $\bm{T}^{c}_{i} = \sum_{i \neq j} a_i \bm{n}_{ij} \times \bm{F}^{\text{tan}}_{ij}$, where $a_i$ is the radius of particle $i$.
The normal force is explicitly expressed as 
\begin{equation}
\bm{F}^{\text{nor}}_{ij} =( k_n \delta_{ij}^{n} -  \zeta^{(n)} u_{ij}^{(n)} ) \bm{n}_{ij}, 
\end{equation}
where $k_n$ is the spring constant, $\delta_{ij}^{n}$ is the normal overlap, $\bm{n}_{ij}$ is the normal unit vector between particles, $u_{ij}^{(n)}$ is the normal velocity difference of the contact point   $u_{ij}^{(n)} = u_i^{(n)}-u_j^{(n)}$, and $ \zeta^{(n)} = \sqrt{m_0 k_n}$ is the damping constant, where $m_0$ is the average mass of the suspended particles.
If the tangential contact force is smaller than a slip criterion, the tangential contact force is represented as 
\begin{equation}
\tilde{\bm{F}}^{\text{tan}}_{ij} = k_t \delta_{ij}^{t} \bm{T}_{ij}, 
\end{equation}
where $k_t$, assumed to be $0.2k_n$, is the tangential spring constant, $\delta_{ij}^{t}$ is the tangential compression and $\bm{T}_{ij}$ is the tangential unit vector at the contact point between particles $i$ and $j$.
We adopt the Coulomb friction rules as
\begin{align}
|\bm{F}_{ij}^{\text{tan}}| &= \mu |\bm{F}_{ij}^{\text{nor}}| \quad \text{if } |\tilde{\bm{F}}^{\text{tan}}_{ij}| \geq \mu |\bm{F}^{\text{nor}}_{ij}|  \quad \text{(slip)}, \\
|\bm{F}_{ij}^{\text{tan}}| &= |\tilde{\bm{F}}^{\text{tan}}_{ij}| \quad  \text{if } |\tilde{\bm{F}}^{\text{tan}}_{ij}| \leq \mu |\bm{F}^{\text{nor}}_{ij}|  \quad \text{(stick)},
\end{align}
whereas $\delta_{ij}^{t}$ is updated each time with relative tangential velocity \cite{luding2008}.

Finally, $\bm{F}_i^r$ is the electrostatic repulsive force, also expressed by pairwise interactions as $\bm{F}_{i}^{\rm r} = \sum_{j \neq i} \bm{F}_{ij}^r$.
The explicit expression of $\bm{F}^{r}_{ij}$ is expressed by the Derjaguin-Landau-Verwey-Overbeek (DLVO) theory \cite{derjaguin1941,verwey1948,israelachvili2011} for the double layer electrostatic force as
\begin{equation} 
\bm{F}^{r}_{ij} =  F_0\exp(-h/\lambda) \bm{n}_{ij}, 
\end{equation}
where $F_0=k_B T \lambda_B\hat{Z}^2(e^{a_{\text{min}}/\lambda_B}/(1+a_{\text{min}}/\lambda_B))^2/h^2$ with the charge number $\hat{Z}$, the Bjerrum length $\lambda_B$ and the Debye-H\"{u}ckel length $\lambda$. Note that $\lambda_B$ can be expressed as $\lambda_B=e^2/(4\pi \epsilon_0 \epsilon_r k_B T)$ where $e$, $\epsilon_0$, $\epsilon_r$, and $k_B$ are the elementary charge, the vacuum permittivity, the dielectric constant, and the Boltzmann constant, respectively \cite{israelachvili2011}.
Here, we adopt the Debye length $\lambda = 0.02a_{\text{min}}$.
Our simulation ignores the Brownian force.
Thus, the electrostatic repulsion force is important to prevent the suspended particles from clustering \cite{pradipto2020,mari2014}.

The equation of motion and torque balance for the impactor with mass $m_I$, velocity $u_I$, moment of inertia $I_I$, and angular velocity $\omega_I$ in the LBM-DEM simulation reads
\begin{equation}\label{eq_of_motion_LBM}
	m_I \frac{d \bm{u}_{I}}{dt}  = \bm{F}^{h}_{I} + \bm{F}^{\rm lub}_{I} + \bm{F}^{c}_{I} + \bm{F}^{g}_{I},
\end{equation}
\begin{equation}	
	I_I \frac{d \bm{\omega}_{I}}{dt} = \bm{T}^{h}_{I} +  \bm{T}^{c}_{I} +  \bm{T}^{\rm lub}_{I}.
\end{equation}
$\bm{F}^{g}_I = - m_I \tilde{g} \bm{\hat{z}}$ is the gravitational force acting on the impactor.
Note that the time dependence of the effective gravitational acceleration acting on the impactor is simply ignored in our analysis.
This is one of the error sources in our analysis.
The contact force $\bm{F}^{c}_{I}$ and torque $\bm{T}^{c}_{I}$, which arise from the interactions with the suspended particles, are also calculated by the DEM.
The lubrication force $\bm{F}^{\rm lub}_{I}$ and torque $\bm{T}^{\rm lub}_{I}$ are also calculated in a similar manner as used in suspended particles.
The long-range hydrodynamic force $\bm{F}^{h}_{I}$ and torque $\bm{T}^{h}_{I}$ are calculated using the bounce-back rule which satisfies the no-slip boundary condition between the fluid and the surface of the impactor \cite{ladd1994a,ladd1994b}.
In the bounce-back rule, the LBM discrete distribution function that streams from fluid nodes to the boundary nodes is reflected.
Then, the hydrodynamic force on each node is calculated from the momentum transferred in this reflection process.
In our implementation, the bounce-back rule is implemented by treating the surface of the impactor as boundary nodes.

\section{\label{app:exp} Comparison with experiments}

\begin{figure}[htbp]
    \centering
	\subfloat[]{\label{fig:7a}%
		\includegraphics[width=0.8\linewidth]{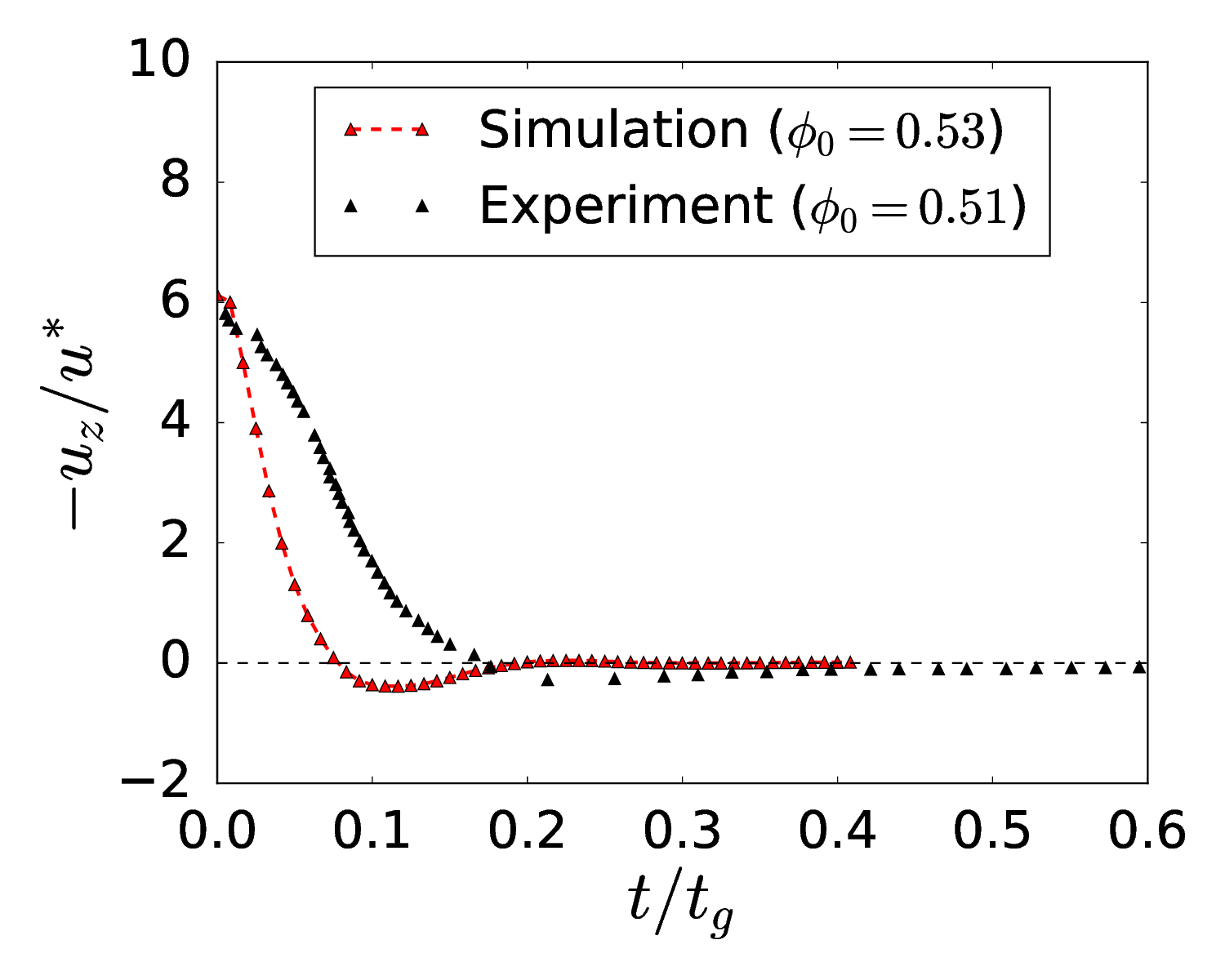}
	}
 
		\subfloat[]{\label{fig:7b}%
		\includegraphics[width=0.8\linewidth]{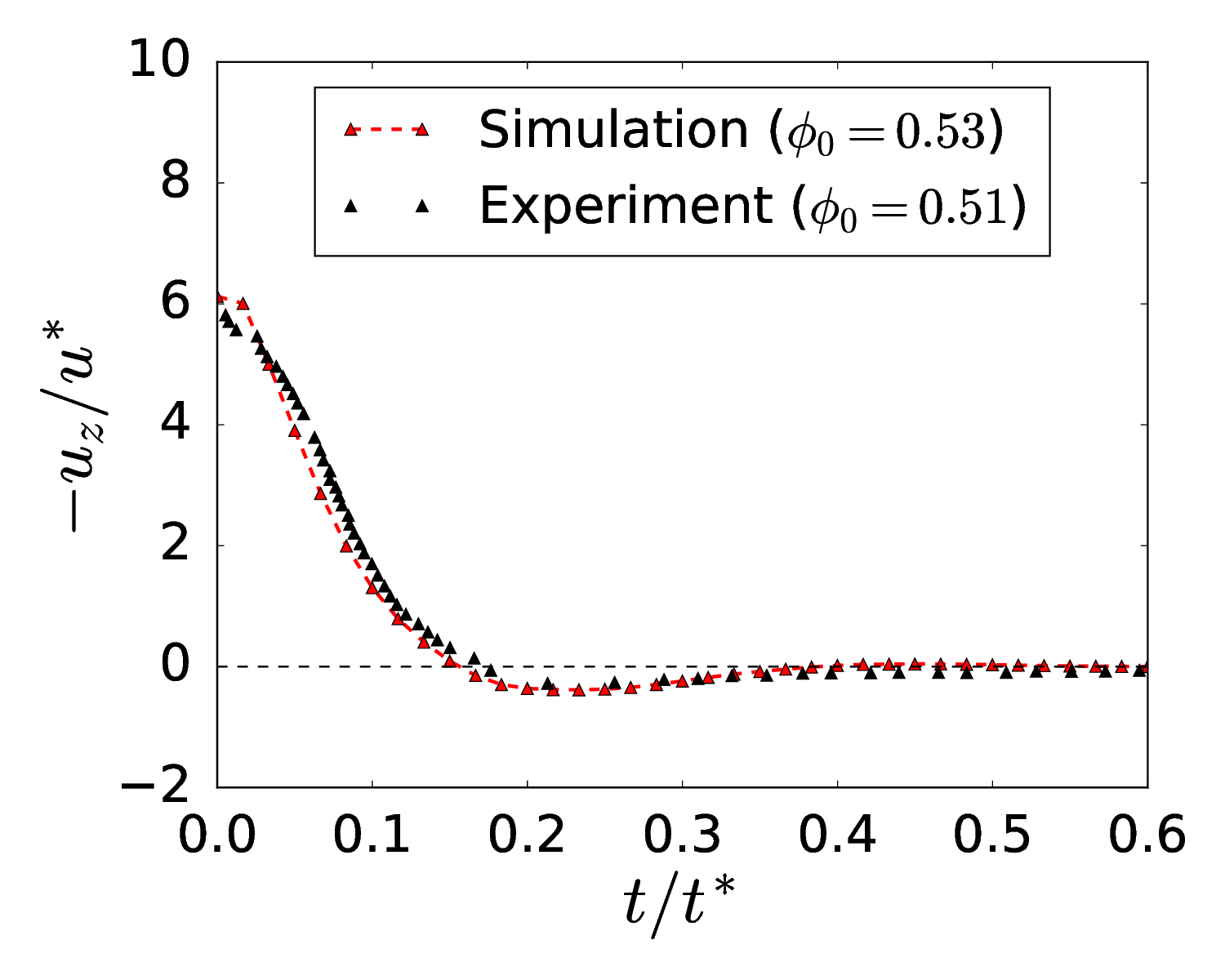}
	}
	\caption{ (a) Plots of the time evolution of the dimensionless impactor velocity for experiment \cite{egawa2019} and LBM-DEM simulation with the scaled time (a) $t_g$ and (b) with $t^{*}$, where $t^{*} = t_g$ for the experiment and $t^{*} = t_g /2$ for the simulation, respectively.
 }
	\label{fig:7}
\end{figure}

\begin{figure*}[htbp]
        \centering
    \includegraphics[width=0.8\linewidth]{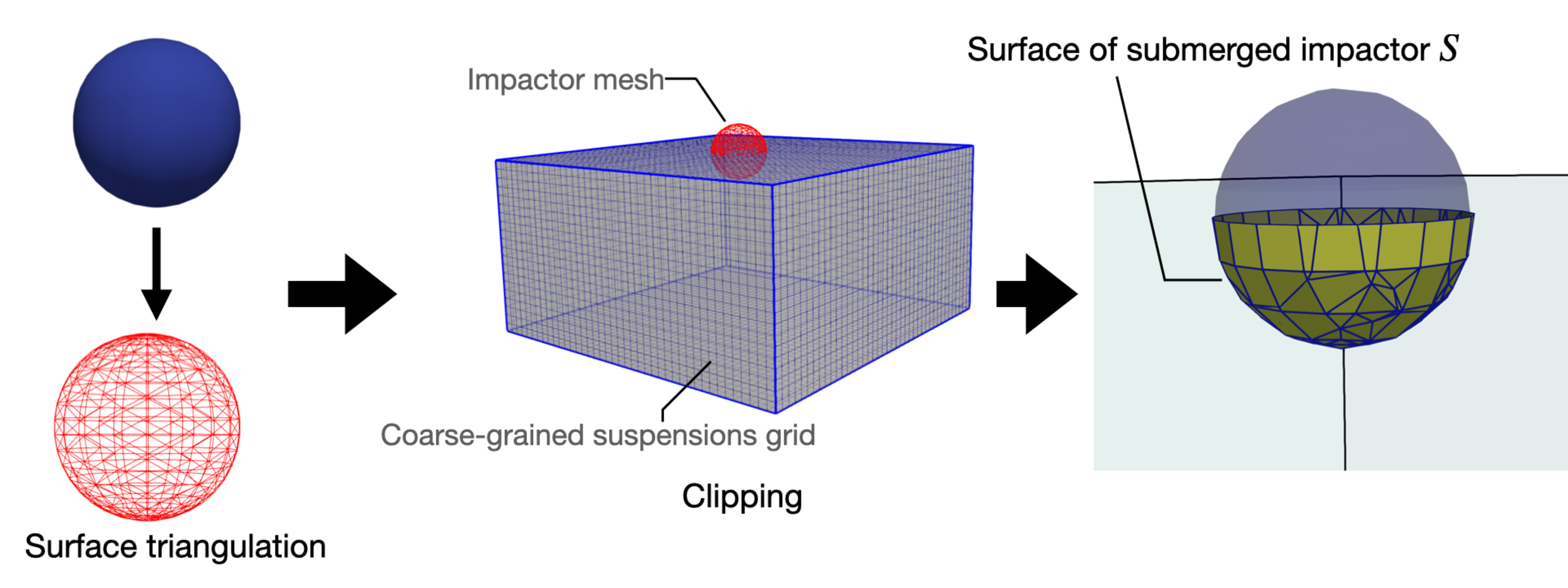}
	\caption{Illustration of the procedure to delineate the surface mesh of the submerged impactor $S$. From left to right: Triangulation of the surface impactor, clipping the impactor surface mesh with the grid of coarse-grained suspensions, resulting surface of the submerged impactor $S$.
 }
	\label{fig:8}
\end{figure*}

In this section, we compare the results of LBM-DEM simulations for the velocity of a free-falling impactor into dense suspensions with a corresponding experiment in the same setup \cite{egawa2019}.
The experimental data is obtained from Fig. 4 in Ref. \cite{egawa2019}, which corresponds to impact velocity $u_0=1.62$ $\rm m/s$, impactor diameter $D_I = 8$ $\rm mm$, impactor density $\rho_I = 8\times10^3$ $\rm kg\:m^{-3}$  and suspensions thickness $H = 20$ $\rm mm$.
In Fig. \ref{fig:7a}, we plot the time evolution of dimensionless velocities $u_z/u^{*}$ for the same dimensionless impact velocity $u_0/u^{*}$. 
Here, one can see that the impact dynamics in the experiment is faster than that in the simulation.
When we introduce another timescale $t^{*}$, which is equal to $t_g$ for the experiment and $t_g /2$ for the simulation, the scaled plot of the simulation perfectly agrees with that of the experiment as shown in Fig. \ref{fig:7b}.

The timescale discrepancy between the simulation and experiment originates from the finite size effect in our simulation, i. e. $t^*\sim N^{0.35}$ as shown in Ref. \cite{pradipto2021}.
Needless to say, our simulation size is much smaller than the experimental counterpart.
We also note that our simplified treatment $F_I^g$ in Eq.~\eqref{eq_of_motion_LBM} is an error source.

In addition to the discrepancy in timescale, the LBM-DEM simulation requires a higher volume fraction than that in the experiments.
This may be from the following:
Our LBM-DEM does not consider the rolling friction, which must exist in the actual consider cornstarch particles.
It is known that the rolling friction lowers the critical volume fraction of discontinuous shear thickening \cite{singh2020}.

\section{\label{app:surf} Evaluation of the surface integrals}

In this section, let us describe the method to delineate the surface of submerged impactor $S$ and perform surface integrals on it.
First, we triangulate the surface of the impactor into a mesh.
The resolution for the triangulation is  30 points in the latitude and the longitude directions.
Then, one can get the submerged surface by clipping the impactor surface mesh with the rectangular grid of the suspensions.
The clipping process is done by keeping the impactor mesh polygons that intersect with the suspensions rectangular grid.
During this clipping process, the polygonal mesh of the surface is persisted.
Finally, the submerged impactor surface mesh $S$ with normals $\bm{n}$ and element $dS$ can be obtained.
Note that each element also contains the variables from the intersecting suspensions grid.
The illustration of this procedure can be seen in Fig. \ref{fig:8}.

Once the submerged surface mesh is obtained, the surface integral can be performed by treating each polygon on the mesh as the integration element. 
Since each polygon in the surface mesh is planar, one can calculate the area of each element and the integration can be done in a straightforward manner (no quadrature required).
These calculations are performed using PyVista, an interface for Visualization Toolkit (VTK) in Python \cite{sullivan2019}.

\section{\label{app:perturbation} Perturbation approach}

In this section, we analytically solve Eq. \eqref{eq:eom} by using a perturbation method in which we assume that the elastic force is much smaller than the viscous force.
Let us introduce the dimensionless depth $\zeta= z/a_I$ and the dimensionless time $\tau :=  3\pi \eta_{\rm eff}a_I t/m_I$.
Thus, the equation of motion is given by
\begin{equation}
    \frac{d^2 \zeta }{d \tau^2} = - G - \frac{d \zeta}{d \tau} \zeta - \epsilon \zeta,
    \label{eq:dim}
\end{equation}
where $G= m_I^2 \tilde{g} / 9 \pi \eta_{\rm eff}^2 a_I^3 $ and $\epsilon= m_I k / (3 \pi \eta_{\rm eff} a_I)^2$.
The perturbation solution of Eq. \eqref{eq:dim} is expressed as
\begin{equation}
    \zeta = \zeta_0 + \epsilon \zeta_1 + O(\epsilon^2) ,
    \label{eq:pert}
\end{equation}
where $\zeta_0$ is the solution of the floating model, i.e. without consideration of elastic force acting on the impactor~\cite{pradipto2021b}.
Plugging Eq. \eqref{eq:pert} into Eq. \eqref{eq:dim}, ignoring higher order terms, and rearranging, one can get up to the first order in $\epsilon$
\begin{align}
    \frac{d^2 \zeta_0 }{d \tau^2} + G +\frac{d \zeta_0}{d \tau} \zeta_0 &= 0, \label{eq:zeta0} \\
       \frac{d^2 \zeta_1 }{d \tau^2}  + \zeta_0 \frac{d \zeta_1}{d \tau}+ \frac{d \zeta_0}{d \tau} \zeta_1 + \zeta_0 &= 0\label{eq:zeta1}
\end{align}
The solution of Eq.~\eqref{eq:zeta0} under the initial conditions $\zeta_0 (0) = 0$ and $ \zeta_{0}^{\prime} (0)  = \tilde{u}_0$, with dimensionless impact velocity $\tilde{u}_0:= u_0 m_I / 3 \pi \eta a_{I}^{2}$, can be written in terms of Airy functions~\cite{pradipto2021b}:
	\begin{equation}
	\zeta_0 (\tau) =\frac{  \kappa[   -   \text{Ai}^{\prime}  (\Phi)\text{Bi}^{\prime} (\Theta)    + \text{Ai}^{\prime} (\Theta)  \text{Bi}^{\prime} (\Phi) ]  } { \gamma [  \text{Bi} (\Phi) \text{Ai}^{\prime} (\Theta)  -  \text{Ai} (\Phi)   \text{Bi}^{\prime} (\Theta)] },
	\label{eq:sol}
\end{equation}
where $\gamma =-G^{2/3}  $, $\kappa = 2^{2/3}G$, $\Phi = (-G\tau + \tilde{u}_0)/(2^{1/3}(-G)^{2/3})$, and $\Theta = ( \tilde{u}_0)/(2^{1/3}(-G)^{2/3}) $.
Here, $\text{Ai} (x)$ is the Airy function of the first kind, which is defined as $\text{Ai} (x) =  \int_{0}^{\infty} \cos ( t^3/3 + xt )dt / \pi $, and $\text{Ai}^{\prime}(x)$ is its derivative. 
$\text{Bi}(x)$ is the Airy function of the second kind, which is defined as $\text{Bi} (x) =  \int_{0}^{\infty}[ \exp (- t^3/3 + xt ) +  \sin (-t^3/3 + xt) ]dt / \pi $, and $\text{Bi}^{\prime}(x)$ is its derivative. 

The solution for Eq. \eqref{eq:zeta1} can be written in terms of the complementary and particular solutions
\begin{equation}
    \zeta_1 (\tau) = \zeta_{1,c} (\tau) +\zeta_{1,p} (\tau). \label{eq:zeta1_cp}
\end{equation}
The complementary solution $ \zeta_{1,c} (\tau) $ can be solved by first solving the homogeneous equation
\begin{equation}
           \frac{d^2 \zeta_1 }{d \tau^2}  +\zeta_0 \frac{d \zeta_1}{d \tau} + \frac{d \zeta_0}{d \tau} \zeta_1  = 0.
           \label{eq:hmgns}
\end{equation}
Then, the complementary solution can be written as
\begin{widetext}
\begin{align}
     \zeta_{1,c} (\tau) &=  c_a\zeta_{a} (\tau) + c_b\zeta_{b} (\tau),\label{eq:zetac} \\
     \zeta_{a} (\tau) &= \frac{2^{\frac{1}{3}} \gamma  \Lambda_1 (\Theta,\Phi) + (Gt - u_0) \Lambda_2(\Theta,\Phi) -2^{\frac{4}{3}} \gamma   \Lambda_3(\Theta,\Phi) + 2^{\frac{1}{3}} \gamma  \Lambda_4(\Theta,\Phi)}{G \left[\text{Ai}^{\prime} (\Theta)\text{Bi} (\Phi) - \text{Ai} (\Phi) \text{Bi}^{\prime} (\Theta)    \right]^2}, \label{eq:zeta}\\
     \Lambda_1 (\Theta,\Phi)&= \text{Ai}^{\prime} (\Phi)^2 \text{Bi}^{\prime} (\Theta)^2, \notag\\
     \Lambda_2 (\Theta,\Phi)&= [\text{Ai}^{\prime} (\Theta) \text{Bi}^{\prime} (\Phi) - \text{Ai} (\Phi) \text{Bi}^{\prime} (\Theta) ]^2, \notag\\
     \Lambda_3 (\Theta,\Phi) &= \text{Ai}^{\prime} (\Theta) \text{Ai}^{\prime} (\Phi) \text{Bi}^{\prime} (\Theta) \text{Bi}^{\prime} (\Phi), \notag\\
     \Lambda_4 (\Theta,\Phi) &=   \text{Ai}^{\prime} (\Theta)^2 \text{Bi}^{\prime} (\Phi)^2, \notag\\
      \zeta_{b} (\tau) &= \frac{1}{\left[\text{Ai}^{\prime} (\Theta)\text{Bi} (\Phi) - \text{Ai} (\Phi) \text{Bi}^{\prime} (\Theta)    \right]^2}, \label{eq:zetb}
\end{align}
\end{widetext}
where $c_a$ and $c_b$ are coefficients that will be determined later from the initial conditions.
Then, the particular solution $\zeta_{1,p}$ can be obtained when we have finite Wronskian, $W (\zeta_a(\tau), \zeta_b(\tau)) \neq 0$ defined as
\begin{equation}
    W (\zeta_a(\tau), \zeta_b(\tau)) := \zeta_{a} (\tau) \zeta_{b}^{\prime}(\tau) - \zeta_{b}(\tau) \zeta_{a}^{\prime}(\tau).
\end{equation}
Plugging Eqs. \eqref{eq:zeta} and \eqref{eq:zetb}, one can obtain
\begin{equation}
    W (\zeta_a(\tau), \zeta_b(\tau)) = -\frac{1}{\left[\text{Ai}^{\prime} (\Theta)\text{Bi} (\Phi) - \text{Ai} (\Phi) \text{Bi}^{\prime} (\Theta)    \right]^2}.
\end{equation}
The particular solution can be written as
\begin{align}
    \zeta_{1,p}(\tau) = \zeta_a (\tau) &\int \frac{\zeta_b (\tau) \zeta_0 (\tau^{\prime})}{  W (\zeta_a(\tau^{\prime}), \zeta_b(\tau^{\prime}))} d\tau \notag\\ &-\zeta_b (\tau) \int \frac{\zeta_a (\tau) \zeta_0 (\tau^{\prime})}{ W (\zeta_a(\tau^{\prime}), \zeta_b(\tau^{\prime}))} d\tau^{\prime},  \\
         =\zeta_a (\tau)& \left[\log \left( \zeta_b (\tau) \right) + C\right] \notag\\ & -\zeta_b (\tau) \int \frac{\zeta_a (\tau) \zeta_0 (\tau^{\prime})}{ W (\zeta_a(\tau^{\prime}), \zeta_b(\tau^{\prime}))} d\tau^{\prime}. \label{eq:zetap}
\end{align}
Note that the integral in the second term of the RHS in Eq. \eqref{eq:zetap} cannot be calculated analytically.
Plugging Eqs. \eqref{eq:zetap} and \eqref{eq:zetac} to Eq. \eqref{eq:zeta1_cp}, one can get
\begin{align}
    \zeta_1 (\tau) = &\zeta_a (\tau) \left[\log(\zeta_b (\tau)) + C_1 \right] \notag\\ &+\zeta_b (\tau) \left[C_2 - \int \frac{\zeta_a (\tau) \zeta_0 (\tau^{\prime})}{ W (\zeta_a(\tau^{\prime}), \zeta_b(\tau^{\prime}))} d\tau^{\prime} \right], \label{eq:zeta1sol}
\end{align}
where $C_1$ and $C_2$ are constants that will be determined from the initial conditions.
Since the integral in the second term in the RHS of Eq. \eqref{eq:zeta1sol} cannot be calculated analytically, a numerical evaluation for this equation is necessary.

Let us discuss the appropriate initial conditions for this perturbation problem.
Note that the perturbative solution only exists ($\epsilon \neq 0$) when elastic force exists after $\tau_{\rm on}$
\begin{equation}
    \zeta (\tau) =
    \begin{cases}
    \zeta_0 (\tau),   \qquad 0 \leq \tau < \tau_{\rm on}, \\
    \zeta_0 (\tau) + \epsilon \zeta_1 (\tau), \qquad \tau \geq \tau_{\rm on}.
    \end{cases}
    \label{eq:pert_case}
\end{equation}
Thus, the initial conditions for Eq. \eqref{eq:dim} are $\zeta (\tau_{\rm on}) = \zeta_{\rm on}$ and $\frac{d \zeta}{d \tau} (\tau_{\rm on})= \tilde{u}_{\rm on}$.
Currently, $\tau_{\rm on} = 3\pi \eta_{\rm eff}a_I t_{\rm on}/m_I$ is another fitting parameter and the value that correspond to $\zeta_{\rm on} =  z_{\rm on} / a_I$ used in Fig. \ref{fig:3b} is chosen.
Then, $\tilde{u}_{\rm on} = u_{\rm on}  m_I / 3 \pi \eta a_{I}^{2}$ can be obtained from the solution of the floating model after specifying $\tau_{\rm on}$.
Then, after ignoring the higher-order terms
\begin{align}
    \zeta_0(\tau_{\rm on}) + \epsilon \zeta_1 (\tau_{\rm on})  &= \zeta_{\rm on} \label{eq:initz}\\
    \frac{d\zeta_0}{d \tau} (\tau_{\rm on})+ \epsilon \frac{d\zeta_1}{d \tau} (\tau_{\rm on})   &=  \tilde{u}_{\rm on}. \label{eq:initu}
\end{align}
Eq. \eqref{eq:initu} must be valid for all $\epsilon$ close to zero.
Nevertheless, assuming some fluctuations in $F_{\rm el}$ one can allow $d \zeta_1(\tau_{\rm on}) / d \tau$ to have some finite but small value $u_{1, \rm on}$.
Therefore, the initial conditions can be written as
\begin{align}
\zeta_0 (\tau_{\rm on}) &= \zeta_{\rm on}, \quad  \zeta_1 (\tau_{\rm on}) = 0, \label{eq:inu} \\
\frac{d \zeta_0}{d\tau} (\tau_{\rm on}) &= \tilde{u}_{\rm on}, \quad  \frac{d \zeta_1}{d\tau} (\tau_{\rm on}) = u_{1, \rm on}.\label{eq:inz}
\end{align}
With this initial condition, $C_1$ and $C_2$ in Eq. \eqref{eq:zeta1sol} can be expressed as
\begin{widetext}
\begin{align}
    C_1 = &\frac{1}{\zeta_a (\tau_{\rm on})}  \Bigg[-\zeta_b (\tau_{\rm on}) \int_{\tau_{\rm on}}^{\tau} \frac{\zeta_a (\tau) \zeta_0 (\tau^{\prime})}{ W (\zeta_a(\tau^{\prime}), \zeta_b(\tau^{\prime}))} d\tau^{\prime}- C_2 \zeta_b (\tau_{\rm on}) -\zeta_a (\tau_{\rm on}) \log \zeta_b (\tau_{\rm on}) \Bigg], \\
    C_2 = &\frac{\zeta_b (\tau_{\rm on}) \left[\frac{\zeta_a (\tau_{\rm on}) \zeta_0 (\tau_{\rm on})}{ W (\zeta_a (\tau_{\rm on}), \zeta_b(\tau_{\rm on}))} + \int_{\tau_{\rm on}}^{\tau} \frac{\zeta_a (\tau) \zeta_0 (\tau^{\prime})}{ W (\zeta_a(\tau^{\prime}), \zeta_b(\tau^{\prime}))} d\tau^{\prime} \right] + \log \zeta_b (\tau_{\rm on}) \left[\zeta_{a}^{\prime} (\tau_{\rm on}) - \zeta_a (\tau_{\rm on})\right]}{\zeta_b (\tau_{\rm on}) - \zeta_{b}^{\prime}(\tau_{\rm on})} \notag\\ +& \frac{\zeta_{b}^{\prime} (\tau_{\rm on}) \left[\frac{\zeta_a (\tau_{\rm on})}{\zeta_b (\tau_{\rm on})} -  \int_{\tau_{\rm on}}^{\tau} \frac{\zeta_a (\tau) \zeta_0 (\tau^{\prime})}{ W (\zeta_a(\tau^{\prime}), \zeta_b(\tau^{\prime}))} d\tau^{\prime} \right] - u_{1,\rm on} }{\zeta_b (\tau_{\rm on}) - \zeta_{b}^{\prime}(\tau_{\rm on}) }, 
\end{align}
\end{widetext}
where $\zeta_{a}^{\prime} = d \zeta_{a} / d\tau$ and $\zeta_{b}^{\prime} = d \zeta_{b} / d\tau$.
Then, we can solve Eq. \eqref{eq:zeta0} with $\tilde{u}_0$ as its initial condition ($\tau = 0$).
Then at $\tau_{\rm on}$ we solve Eq. \eqref{eq:pert} with Eqs. \eqref{eq:inu} and \eqref{eq:inz} as initial conditions and stitch this with the solution of Eq. \eqref{eq:zeta0}. 

\bibliography{references}

\end{document}